\documentclass[aps,pra,twocolumn,longbibliography,groupedaddress,showpacs,floatfix,superscriptaddress]{revtex4-1}
\usepackage{graphicx}  
\usepackage{dcolumn}   
\usepackage[english]{babel}
\usepackage{bm}       
\usepackage{amsmath, amsthm, amssymb}
\usepackage{bbold}
\usepackage{amsmath}
\usepackage{mathrsfs}
\usepackage{array}

\usepackage[dvipsnames,usenames]{color}
\usepackage{soul} 
\usepackage{ulem} 

\usepackage{hyperref}
\hypersetup{
    citecolor=magenta,
    colorlinks=true,
    linkcolor=blue,
    filecolor=magenta,      
    urlcolor=blue,
    pdftitle={Overleaf Example},
    pdfpagemode=FullScreen,
    }

\emergencystretch=\maxdimen
\newcommand{\calH}{\mathcal{H}}

\def\ave#1{\langle #1\rangle}

\begin{document}

\normalem

\title{Quantum percolation on Lieb Lattices}
\author{W. S. Oliveira}
\affiliation{Instituto de F\'\i sica, Universidade Federal do Rio de Janeiro Cx.P. 68.528, 21941-972 Rio de Janeiro, Rio de Janeiro, Brazil}
\author{J. Pimentel de Lima}
\affiliation{Departamento de F\'\i sica, Universidade Federal do Piau\'\i, 64049-550 Teresina, Piau\'\i, Brazil}
\author{Raimundo R. dos Santos}
\affiliation{Instituto de F\'\i sica, Universidade Federal do Rio de Janeiro Cx.P. 68.528, 21941-972 Rio de Janeiro, Rio de Janeiro, Brazil}

\begin{abstract}
We theoretically investigate the quantum percolation problem on Lieb lattices in two and three dimensions. 
We study the statistics of the energy levels through random matrix theory, and determine the level spacing distributions, which, with the aid of finite-size scaling theory, allows us to obtain accurate estimates for site- and bond percolation thresholds and critical exponents. 
Our numerical investigation supports a localized-delocalized transition at finite threshold, which decreases as the average coordination number increases. 
The precise determination of the localization length exponent enables us to claim that quantum site- and bond-percolation problems on Lieb lattices belong to the same universality class, with $\nu$ decreasing with lattice dimensionality, $d$, similarly to the classical percolation problem. 
In addition, we verify that,  in three dimensions, quantum percolation on Lieb lattices belongs to the same universality class as the Anderson impurity model.
\end{abstract}

\date{Version 3.24 -- \today}

\maketitle
\section{Introduction}

The problem of electron diffusion in random media has been an active area of research for over 60 years, since Anderson's pioneering work \cite{Anderson58,Lagendijk09}. 
In what has become known as the Anderson impurity model (AIM), tight-binding electrons move on a lattice with independent random site energies; this is usually referred to as \emph{diagonal} disorder.
One of the main issues has been to establish the conditions under which electrons do not diffuse throughout the lattice, thus becoming localized.  
For sufficiently strong disorder all states are expected to be localized, but, in addition, in one- and two spatial dimensions any amount of disorder leads to localization, as predicted by a scaling theory \cite{Abrahams79}. 
It is interesting to note that, more recently, Anderson localization crossed over the boundaries of matter waves, becoming a generic wave phenomenon in random media \cite{Lagendijk09}. 

A closely related situation is when hopping is `switched off' between randomly chosen pairs of neighboring sites; this may occur because either one of the sites is absent, or an interstitial defect prevents hopping to take place. 
Since this \emph{off-diagonal} disorder immediately connects with the purely geometrical classical percolation problem~\cite{Stauffer94}, the electronic problem is referred to as quantum percolation (QP). 
Despite their similarities, diagonal and off-diagonal disorder differ in many fundamental aspects of localization. 
For example, in two-dimensional geometries there is a consensus that any amount of diagonal disorder leads to localization, but the situation for QP has not yet been fully settled  \cite{Shapir82,Soukoulis91,Berkovits96,Nazareno02,Islam08,Schubert08,Gong09,Dillon14}.
In three dimensions both the Anderson model and the QP model undergo a localization transition at finite disorder; however, while the localized region in the former is bounded by two mobility edges around the band center, in the QP model there is no localization at the band center \cite{Thouless72,Mano2017}.

\begin{figure*}
\centering
\includegraphics[scale=0.37]{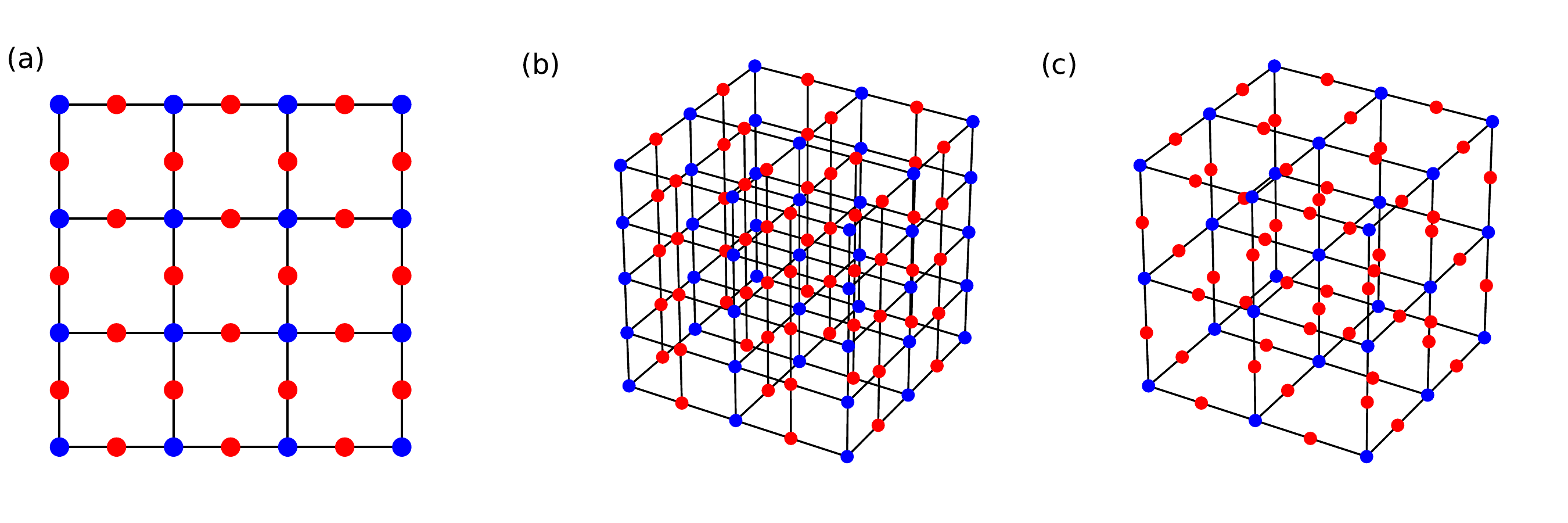}
\caption{The two-dimensional Lieb lattice (a), and its three-dimensional extensions: (b) the layered Lieb lattice, LLL, and (c) the perovskite lattice, PL.
}
\label{fig:lattices}
\end{figure*}

On the other hand, when properly engineered, the interplay between lattice geometry and electronic band structure may give rise to exotic phenomena. 
For instance, the presence of flat (or dispersionless) bands has recently attracted a lot of attention since they may lead to ferrimagnetism \cite{Lieb89,Costa16,Costa18,Yarmohammadi18,Bouzerar23}, topological states \cite{Weeks10,Jiang19,Jiang20,Poata23} and enhanced electronic correlations \cite{Andrade23}.
Several electronic properties of these systems stem from the fact that flat band states are localized due to their high degeneracy.
In two dimensions, for instance, flat bands appear in the so-called Lieb lattices (LL's), also known as CuO$_2$ lattices, where the four-coordinated sites represent Cu atoms and doubly-coordinated sites represent O sites; see Fig.\,\ref{fig:lattices}(a). 
Flat bands may also arise when LL's are piled up along the direction perpendicular to the CuO$_2$ planes.
However, one can either pile up layers of 2D LL's, as shown in Fig.\,\ref{fig:lattices}(b), or one can form a perovskite lattice (PL), in which an `O'  site is introduced along the $c$-axis halfway between two CuO$_2$ layers, so that each face of the cube looks the same, as in Fig.\,\ref{fig:lattices}(c). 
From the point of view of electronic properties, we note that the PL shares the same symmetry of the 2D LL along the three cartesian directions, thus preserving the flat band in the 3D extension \cite{Weeks10}; by contrast, the layered Lieb lattice (LLL) does not display a flat band~\cite{Noda15}, since there is a delocalizing channel through sites along the $c$-axis.
Further, the possibility of studying these geometries using ultracold fermionic atoms in optical lattices has triggered even more interest, due to the exceptional experimental precision and flexibility in adjusting parameters for these systems, including interactions, particle density and disorder.

Since positional disorder may break the degeneracy associated with flat bands, a study of localization in these geometries is certainly of interest.
Indeed, transfer matrix studies suggest that any amount of diagonal (Anderson) disorder leads to localization in a two-dimensional LL~\cite{Liu21}, while for the three-dimensional PL extended states are possible for some amount of disorder~\cite{Liu20,Liu21}.
We should have in mind that the amount of disorder in the AIM is measured through electronic parameters such as the width of the distribution of site energies; by contrast, disorder in QP is measured geometrically, through the concentration, $p$, of active sites (or bonds). 
This allows us to gain considerable insight by comparing quantum percolation thresholds for different lattices, especially to probe how geometries with flat bands respond to disorder.
Indeed, due to quantum interference, localization can occur even when a spanning cluster is present, so one expects that quantum percolation thresholds are larger than the classical ones, i.e., $p_{q} > p_{c}$; for a discussion of classical percolation on LL's, see Ref.\,\cite{Oliveira21}. 
With this in mind, here we study some critical properties for the QP model on 2D and 3D Lieb lattices through numerical calculations of the level statistics of the system. 
Level statistics, rooted in Random Matrix Theory (RMT), is a useful tool to probe the underlying physics of disordered quantum systems \cite{Andreev96, Muttalib87, Beenakker97,wigner57}. 
For instance, a single quantum particle interacting with random impurities in the delocalized regime exhibits correlations between the eigenvalues consistent with Gaussian matrix ensemble predictions; by contrast, in the localized regime, these eigenvalue correlations vanish, resulting in poissonian level statistics. 
However, at the precise point of the localized-delocalized transition, the level statistics deviate from those predicted by Gaussian matrix ensembles \cite{Shklovskii93, Kottos02, Zharekeshev95, Nishigaki99}, thus allowing one to determine some of the critical properties of disordered systems; examples of its use in the AIM and QP on 3D lattices can be found in Refs.\,\cite{Berkovits96,Travenec08,Nishigaki99}.

The layout of the paper is as follows: In Sec.\,\ref{sec:HQMC} we present the Hamiltonian and briefly describe the way disorder configurations are generated, how the level spacing distribution is calculated, and the finite-size scaling method used to obtain critical concentrations and critical exponents. 
In Sec.\,\ref{sec:results} we present and discuss the results obtained for Lieb lattices, and Sec.\ref{sec:conc} summarizes our findings.

\section{Model and Method}
\label{sec:HQMC}

Quantum percolation is usually formulated in terms of a tight-binding Hamiltonian for a single spinless electron,
\begin{align}
\label{eq:Ham}
	\calH= \sum_{i} \varepsilon_{i} a_{i}^\dagger a_{i}^{\phantom{\dagger}} - \sum_{\ave{i,j}}\left(t_{ij}a_{i}^\dagger 
a_{j}^{\phantom{\dagger}}+\mathrm{H.c.}\right),
\end{align}
in which $\varepsilon_{i}$ is the site energy, $a_{i}^\dagger$ and $a_{j}$ are electron creation and annihilation operators respectively, $t_{ij}$ is the hopping energy and $\ave{i,j}$ denote nearest neighbors. 
In the quantum site percolation problem we consider a lattice whose sites are occupied at random and independently with probability $p$, and empty with probability $q =1 - p$; thus, $t_{ij}=1$ if sites $i$ and $j$ are occupied, and $t_{ij}=0$ otherwise. 
In the quantum bond percolation problem, the hopping matrix elements $t_{ij}$ are randomly taken as $1$ or $0$ with probabilities $p$ and $q=1 - p$, respectively. 
Since we are dealing with off-diagonal disorder, we set the onsite energy as a constant, which we take as  $\varepsilon_{i} = 0, \,\forall i$, without loss of generality.  

\begin{figure}[t]
\centering
\includegraphics[scale=0.47]{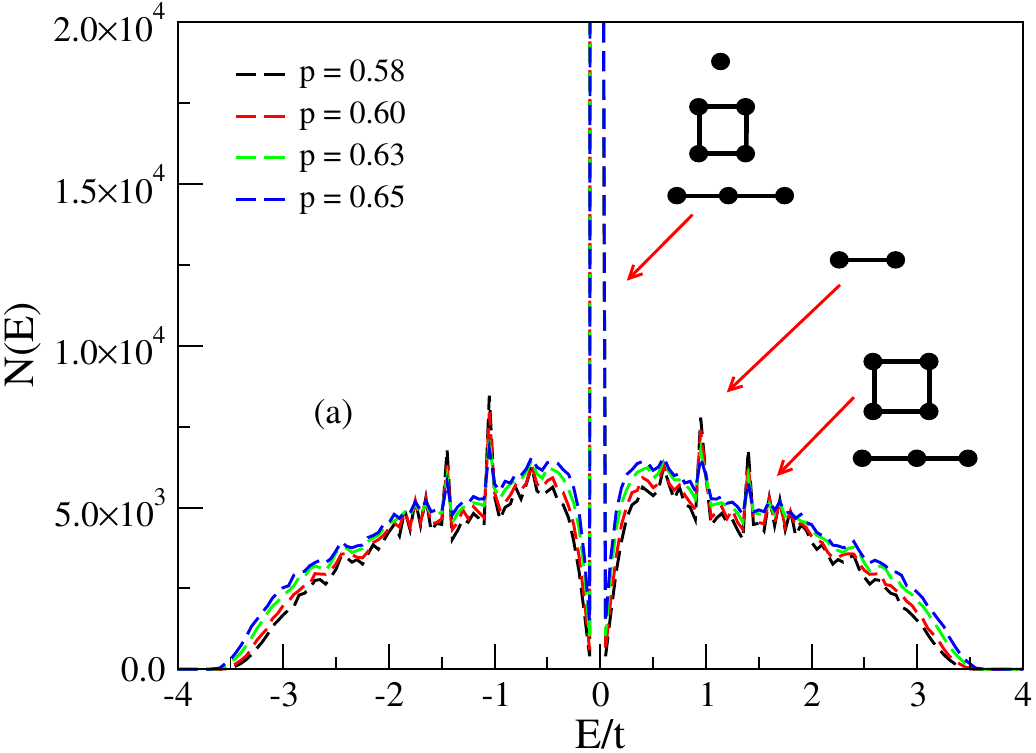}
\includegraphics[scale=0.47]{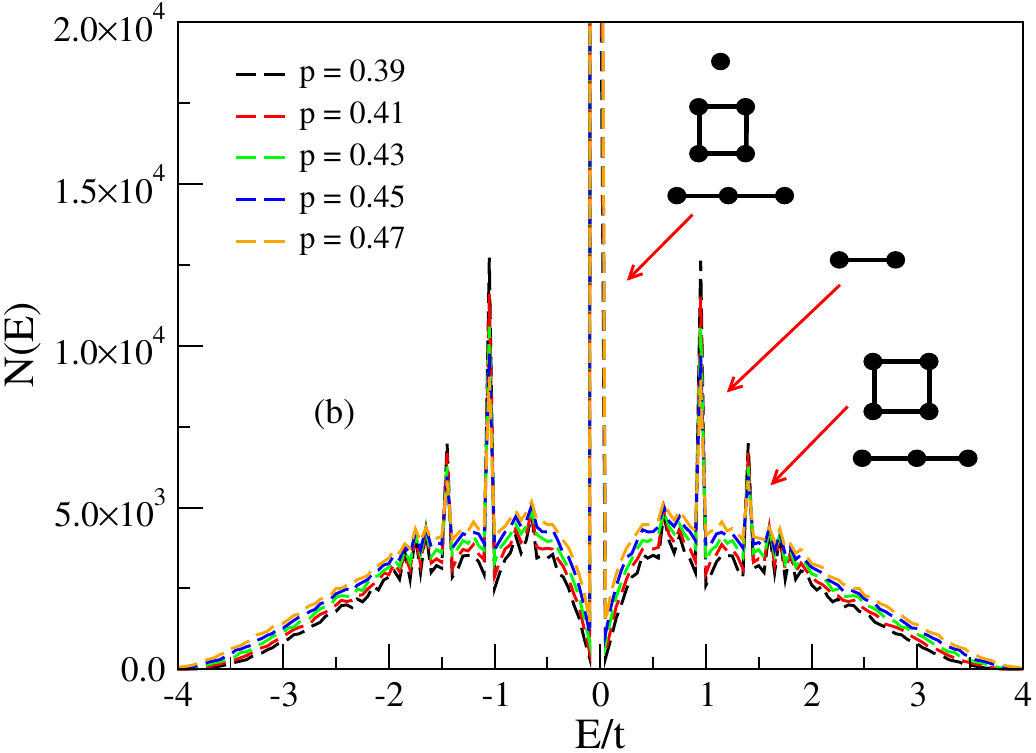}
\caption{Density of states, $N(E)$,  for different site occupation probabilities for (a) square,  and (b) simple cubic lattices, with linear sizes $L = 34$ and $L = 15$, respectively; results for bond disorder are similar. 
The peaks are contributions from small isolated clusters; see text.
}
\label{fig:DOS}
\end{figure}

In order to test the accuracy of our method, we first analyze the level statistics for quantum site percolation on the square and simple cubic lattices. 
We consider lattices with linear size, $L$, and periodic boundary conditions, and, for a given disorder configuration with a fraction $1-p$ of sites rendered inactive, we diagonalize the Hamiltonian,  Eq.\,\eqref{eq:Ham}, to obtain its eigenvalues. 
This procedure is then repeated for different occupation probabilities to yield the density of states (DOS) displayed in Fig.\,\ref{fig:DOS}. 
For both lattices, the most visible feature is the emergence of a series of sharp peaks which increase as $p$ decreases. These peaks are attributed to the formation of small disconnected clusters of sites~\cite{Berkovits96,Cao12}. 
For instance, the peaks around $E/t=\pm 1$ are contributions from isolated clusters with two sites, while the central peak is due to single sites, as well as to clusters with three and four sites; note that clusters with three and four sites also contribute to the peaks near $E/t=\pm 1.4$, depending of the shape of the cluster and if all its sites are connected or not.  
In order to mitigate the contribution from these isolated clusters, in what follows we will only include in our tight-binding model the sites belonging to the percolating cluster. 
Although this restricts us to work with an occupancy probability above the classical percolation threshold, we stress that the quantum percolation thresholds are expected to be greater than, or equal to, the corresponding classical ones. 
We resort to the Hoshen-Kopelman algorithm~\cite{Hoshen76} to select these working percolating clusters. 
Note that the issue of uniqueness of a percolating cluster~\cite{Haggstrom06} is immaterial in the present case, since our concern here is with localization in a single percolating cluster.

\begin{figure}[t] 
\centering
\includegraphics[width=8.0cm]{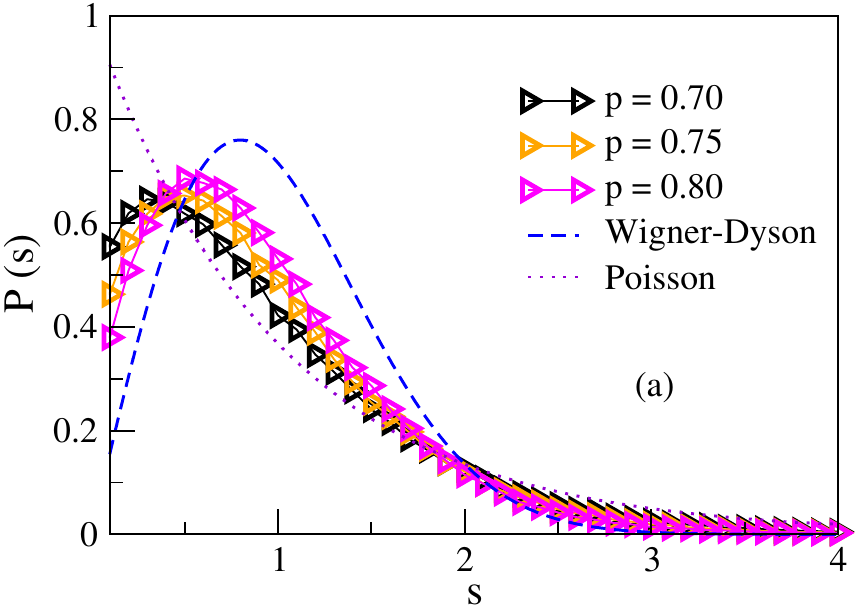}
\includegraphics[width=8.0cm]{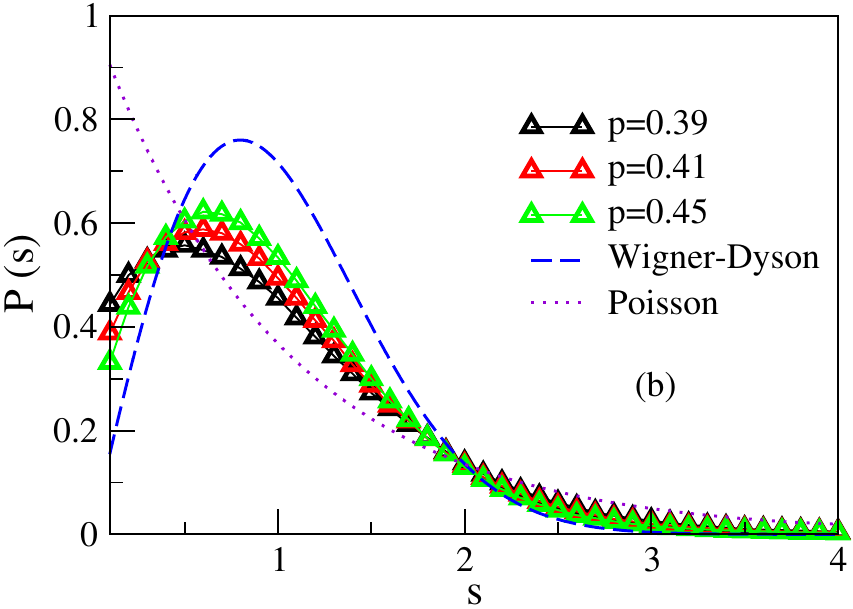}
\caption{The level spacing distribution for different concentrations of active sites for the square (a) and the simple cubic (b) lattices, respectively with linear sizes $L = 34$ and $L = 15$ . 
For comparison, we also show analytical expressions for the gaussian orthogonal ensemble distribution (GOE) and the Poisson distribution. 
}
\label{fig:distributions}
\end{figure}

\begin{figure*}
\centering
\includegraphics[scale=0.65]{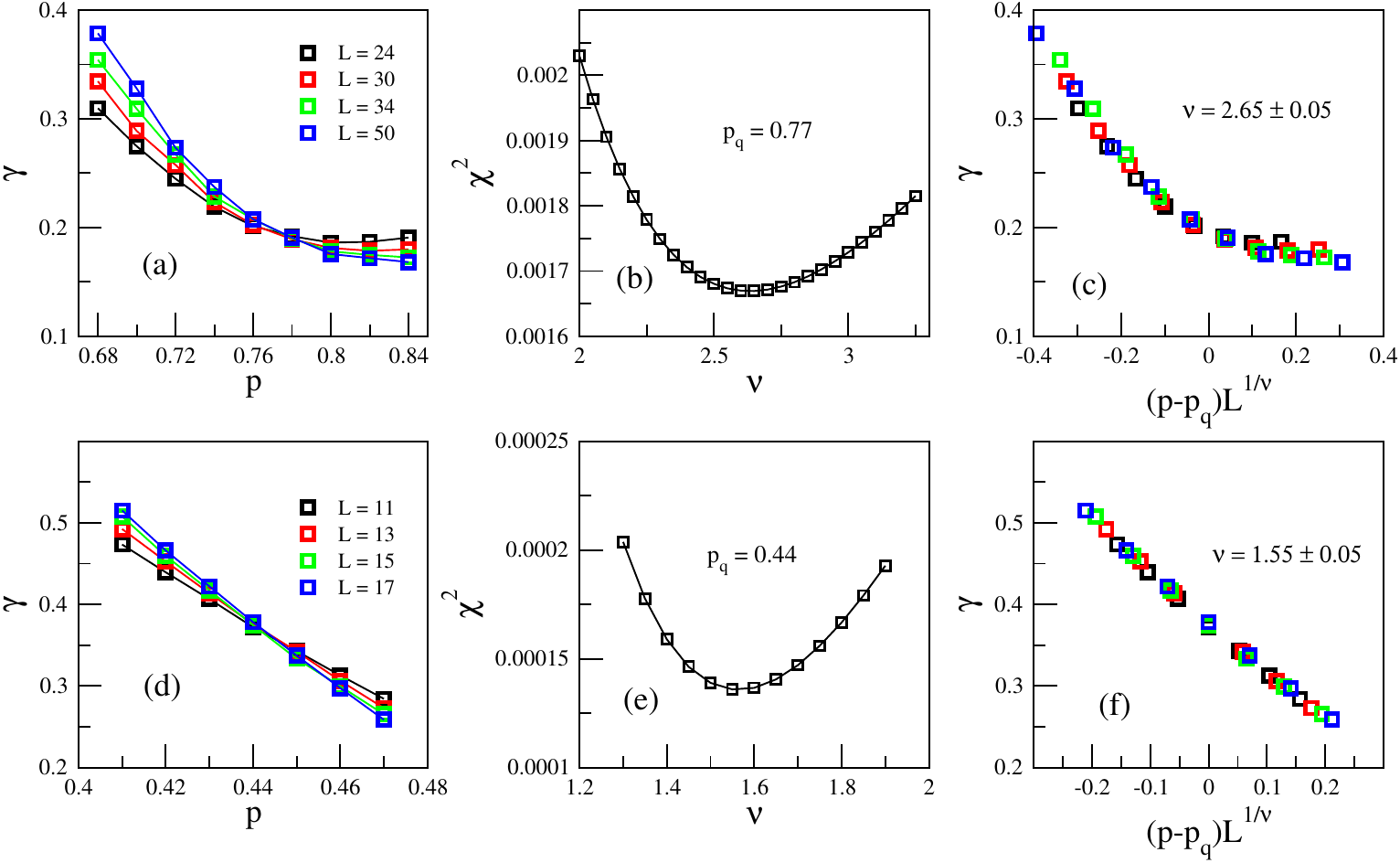}
\caption{The function $\gamma (p,L)$, for different system sizes and occupation probabilities, for the square\,(a) and simple cubic\,(d) lattices on the quantum site percolation. The chi-squared values of a polynomial fit to the data collapse is shown in (b) for the square lattice and (e) for the simple cubic lattice, for fixed $p_q$.  The optimal data collapse is shown in (c) and (f) for the square and simple cubic lattice respectively.}
\label{fig:gamma_square_and_cubic}
\end{figure*} 

For a given disorder configuration, the classical percolating cluster gives rise to, say $K+1$ energy eigenvalues forming a bandwidth, $W$, so that the average level spacing is $W/K$.
The latter is taken to set the scale of level spacings through  
\begin{equation}	
	s_i \equiv \frac{E_{i+1}-E_i}{W/K}.
\end{equation}
We generate data for $M$ disorder configurations, which for the square lattice varies between 400 and 1,800, respectively for $L=50$ and $L=24$; for the simple cubic lattice $M$ varies between 200 and 750, respectively for  $L=17$ and $L=11$; this ensures meaningful statistics of about $10^6$ eigenvalues per lattice size.

At this point it is crucial to distinguish between two formulations of the QP problem. 
One may ask either if the particle diffuses irrespective of its energy, or if it diffuses when it has an energy within a given interval. 
Here we will be concerned solely with the former. Nonetheless, for completeness we should mention that in two dimensions the latter formulation seems to give rise to a two-step localization-delocalization transition as $p$ increases: first, from an exponentially localized to power-law localized, and then to delocalized \cite{Islam08,Dillon14}.
This feature is absent for a simple cubic lattice.

Figure \ref{fig:distributions} shows the corresponding level spacing distributions, $P(s)$, for three different concentrations, $p$; note, however, that the concentrations considered in panel (a) are centered around values larger than those in panel (b), simply by virtue of the fact that the classical $p_c$ decreases with the dimension of the hypercubic lattices.
Also shown are two expected limiting cases of the distribution: the Poisson behavior, $P(s)=e^{-s}$, characteristic of the localized regime \cite{Mirlin00}, and the Wigner-Dyson behavior, $P(s) =(\pi/2)\,s\,e^{\pi s^2\!/4}$, as obtained from the Gaussian Orthogonal Ensemble (GOE)  \cite{Mirlin00,Altshuler86}.
As $p$ increases, we see that in both cases  $P(s)$ decreases in the small $s$ region, while it increases above $s\approx 0.5$. 
In addition, one should also note that all curves seem to intersect at $s \approx 2$, which is the same feature observed for the Anderson model in three dimensions \cite{Shklovskii93}. 
The inescapable conclusion is that, as $p$ increases the distribution evolves from Poisson to Wigner-Dyson, that is, from localized to delocalized. 

In order to obtain an accurate estimate for $p_q$ in these cases, we calculate the weight of the large-$s$ area in $P(s)$, that is, beyond  the crossing point at $s=2$ \cite{Berkovits96,Kaneko99,Shklovskii93,Cao12},
\begin{equation}
	A=\int_{2}^{\infty} P(s) \,ds,
\label{eq:A}
\end{equation}
normalized through
\begin{equation}
	\gamma(p,L) = \frac{A - A_{W}}{A_{p} - A_{W}},
\label{Eq. gamma_parameter}
\end{equation}
where
\begin{equation}
	A_\text{WD}\equiv \int_{2}^{\infty} \frac{\pi s}{2} e^{\pi s^2/4} \,ds = e^{-\pi}
\end{equation}
 and 
 \begin{equation}
 A_\text{P}\equiv\int_{2}^{\infty} e^{-s} \,ds = e^{-2}
\end{equation}
are the large-$s$ weights of the Wigner-Dyson and Poison distributions, respectively.

Figures \ref{fig:gamma_square_and_cubic}(a) and (d) respectively show plots of $\gamma (p,L)$ for the square and simple cubic lattices for various systems sizes. 
The behavior noted in Fig.\,\ref{fig:distributions} is more clearly identified here, namely a regime in which $\gamma (p,L)$ increases with increasing $L$, and another in which it decreases with $L$.
Separating these regimes there is a concentration, $p_q$, at which $\gamma (p_q,L)$ is size-scaling invariant: $p_q^\text{sq}\approx 0.77$ and $p_q^\text{sc}\approx 0.44$, for the square and simple cubic lattices, respectively.
One therefore expects that $\gamma(p,L)$ follows a finite-size scaling \textit{ansatz} \cite{Fisher71,Barber83}, \begin{equation}
    \gamma (p,L) = f [L/\xi (p)],
    \label{eq. correlationlenght}
\end{equation}
where $\xi (p)$ is the correlation length, which near the critical point behaves as
\begin{equation}
	\xi (p) = \xi_{0} \left| \frac{p}{p_{q}} -1\right|^{-\nu}; 
\end{equation}
note that since $\gamma$ is a dimensionless quantity, there is no $L$ dependence multiplying the scaling function $f[L/\xi (p)]$.
Near $p_q$ we may therefore expect that \cite{Fisher71,Barber83,dosSantos81a} 
\begin{equation}
\label{Eq.gamma_scaling}
	\gamma(p,L) = \gamma(p_{q})+C\left|\frac{p}{p_{q}}-1\right|L^{\frac{1}{\nu}},
\end{equation}
where $C$ is a constant, which allows us to determine the critical exponent, $\nu$, by fitting the data near the crossing point. 
Figures \ref{fig:gamma_square_and_cubic} (c) and (f) show the optimal data collapse for the data appearing in panels (a) and (d), for the square and simple cubic lattice, respectively, in which $\nu$ is considered as an independent variable, adjusted through a least squares fit for fixed $p_q$, as displayed in panels (b) and (e).

\begin{table*}[t]
\centering
\begin{tabular}{l|l|l|l|l|l}
\hline
Lattice &       & site              & bond              & $\nu_q$        &Comments     \\ 
\hline
\hline
Square  & $p_c$ & $0.59274598(4)^a$ &   $1/2^b$          &  $0.57(1)^c$   &$p_q^{\text{(s)}} = 0.740(25)^c$, $p_q^{\text{(b)}} = 0.625(25)^c$, $p_q^{\text{(b)}} = 0.60(4)^m$\\ 
\cline{2-4}
        & $p_q$ & $0.75(1)^l$       & $0.65(1)^l$        &   $2.65(5)^l$, $3.34^n$, $[2.6-3.2]^o$  &     \\ 
\hline
LL      & $p_c$ & $0.7396(5)^j$     & $0.6438(3)^j$       & $2.55(5)^l$, $2.60(5)^l$    &    \\ 
\cline{2-4}
        & $p_q$ & $0.91(1)^l$       & $0.82(1)^l$         &                 &    \\ 
\hline
SC      & $p_c$ & $0.311681(13)^d$  & $0.24881182(10)^d$  &                  &    \\ 
\cline{2-4}
        & $p_q$ & $0.44(1)^{l}$ & $0.33(1)^{l}$           &  $1.59(5)^{l}$, $1.58(5)^{f,h,i}$  & $p_q^{\text{(s)}} = 0.44(1)^{e,f,g}$, $p_q^{\text{(b)}} = 0.33(1)^{h}$, $p_q^{\text{(b)}} = 0.33(6)^{m}$   \\ 
\hline
LLL     & $p_c$ & $0.3919(5)^j$           & $0.3338(5)^j$      &  $1.57(5)^{l}$, $1.55(5)^{l}$          &      \\ 
\cline{2-4}
        & $p_q$ & $0.55(1)^l$                & $0.39(1)^l$      &                    \\ 
\hline
 PL     & $p_c$ &  $0.5225(5)^j$           &  $0.4010(5)^j$      &    $1.58(5)^{l}$, $1.55(5)^{l}$        \\ 
\cline{2-4}
        &$p_q$  & $0.71(1)^l$                & $0.62(1)^l$       &            \\
\hline
\end{tabular}
\caption{Estimates obtained for the quantum critical concentration, $p_q$, and for the correlation length exponent, $\nu$; see text.
LL, LLL and PL respectively stand for Lieb lattice, layered Lieb lattice and perovskite lattice; see Fig.\,\ref{fig:lattices}. 
Notes: $^a$Ref.\,\cite{Lee08}; $^b$Ref.\,\cite{stauffer18}; $^c$Ref.\,\cite{Daboul00}; $^d$Ref.\,\cite{Wang13}; $^e$Ref.\,\cite{Soukoulis92}; $^f$Ref.\,\cite{Stadler96}; $^g$Ref.\,\cite{Koslowski90}; $^h$Ref.\,\cite{Ujfalusi14}; $^i$Ref.\,\cite{Kaneko99};  $^j$Ref.\,\cite{Oliveira21}, $^l$This work, $^m$Ref.\,\cite{Meir89}, $^n$Ref.\,\cite{Odagaki84}, $^o$Ref.\,\cite{Gong09}.}
\label{tab:pcnu}
\end{table*}

We have repeated the above procedure for bond QP, and Table \ref{tab:pcnu} shows our estimates for $p_{q}$ and $\nu$, for the quantum site- and bond-percolation on the square and simple cubic lattices; we also show previous estimates for comparison.
Our estimates for the thresholds are in very good agreement with those obtained from different methods.
The same holds for the critical exponent $\nu$, in which case the discrepancy with respect to the value obtained in Ref.\,\cite{Daboul00} may be attributed to the short series used in their expansion; the discrepancy with respect to the estimate from Ref.\,\cite{Odagaki84} is due to the smallness of the cells used in their real-space renormalization group approach;  
These results confirm the reliability of our method, to the point of contributing to add credence to expectation that the QP problem belongs to the same universality class as the AIM \cite{Stadler96,Kaneko99,Berkovits96,Ujfalusi14,Travenec08,Soukoulis92, Stadler96,Koslowski90,Dodoo13,Lee93}.
We now consider the Lieb lattices in turn.


\section{Results for the Lieb lattices}
\label{sec:results}

\subsection{The 2D Lieb lattice}
\label{ssec:2D}

The unit cell for the two-dimensional Lieb lattice contains 3 sites; see Fig.\,\ref{fig:lattices}(a).
Therefore, each cartesian direction of linear size $L$ actually comprises of $2L$ sites; we recall that while the computational effort is proportional to $N_s\equiv 4L^2$, the finite-size scaling parameter is simply $L$. 
The number of disorder realizations, $M$, used was $M = 2000\,(500)$ runs for the smallest (largest) lattice size, $L = 24\,(50)$.

\begin{figure}[t]
\centering
\includegraphics[scale=0.45]{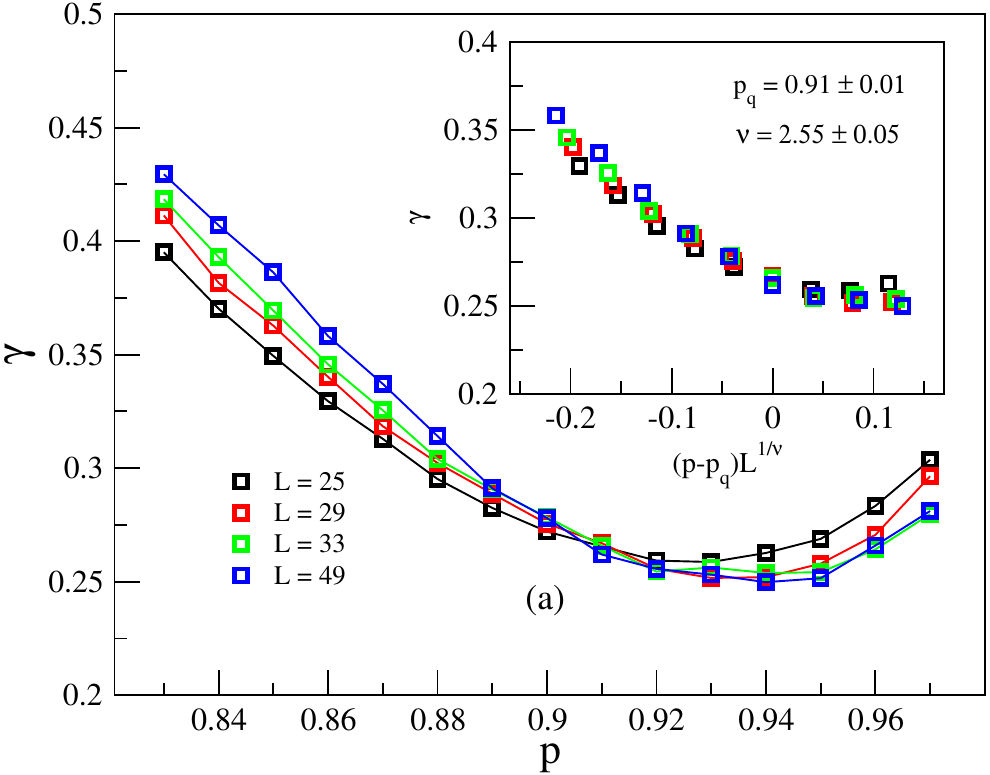}
\includegraphics[scale=0.45]{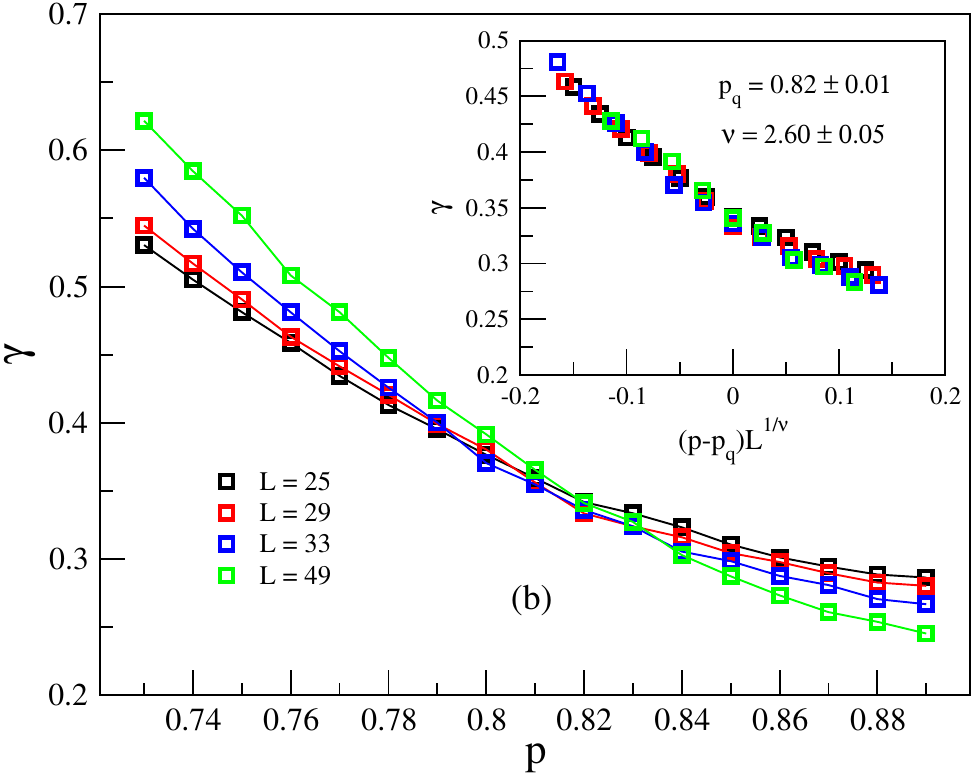}
\caption{The function $\gamma (p,L)$ [Eq.\,\eqref{Eq.gamma_scaling}] for the two-dimensional Lieb lattice, and for different system sizes:  (a)~site and (b) bond QP.
The insets show the data collapse according to Eq.\,\eqref{eq. correlationlenght}}
\label{fig:gamma_lieb2D}
\end{figure}

We characterize the level statistics through the same procedure outlined in Sec.\,\ref{sec:HQMC}.
Figure \ref{fig:gamma_lieb2D} shows $\gamma(p,L)$, Eq.\,\eqref{Eq. gamma_parameter}, as a function of $p$ for different system sizes.
For both bond and site quantum percolation, there is a common crossing point, which is identified as the critical concentration.
Close to the critical point, the size dependence is governed by Eqs.\,\eqref{eq. correlationlenght} and \eqref{Eq.gamma_scaling}, and we adopt the same procedure as for the square and simple cubic lattices; that is, we optimize the data collapse to obtain the critical parameters $p_{q}^{(\text{s})} = 0.91 \pm 0.01$ with $\nu = 2.55 \pm 0.05$  and $p_{q}^{(\text{b})} = 0.82 \pm 0.01$ with $\nu = 2.60 \pm 0.05$, where the superscripts (s) and (b) respectively denote site and bond cases. 
As expected, we see that $p_{q}^{(\text{s})}> p_{q}^{(\text{b})}$, and each is larger than than the corresponding classical ones, namely  $p_{c}^{(\text{s})}=0.7396(5)$ and $p_{c}^{(\text{b})}=0.6438(3)$~\cite{Oliveira21}.
Further, the equality (within error bars) of critical exponents indicates that site and bond QP belong to the same universality class, which is the same as for the square lattice. 
For completeness, we note that the estimates for classical percolation on the Lieb lattice are  $\nu^{(\text{s,\,cl.})}=1.35(4)$ and $\nu^{(\text{b,\,cl.})}=1.30(5)$~\cite{Oliveira21}; these values, in turn, are also consistent with the geometry not changing the universality class.
And, finally, it is worth mentioning that the critical exponent satisfies the inequality $\nu \geq 2/d$ \cite{Chayes86}.

\begin{figure}[h]
\centering
\includegraphics[scale=0.44]{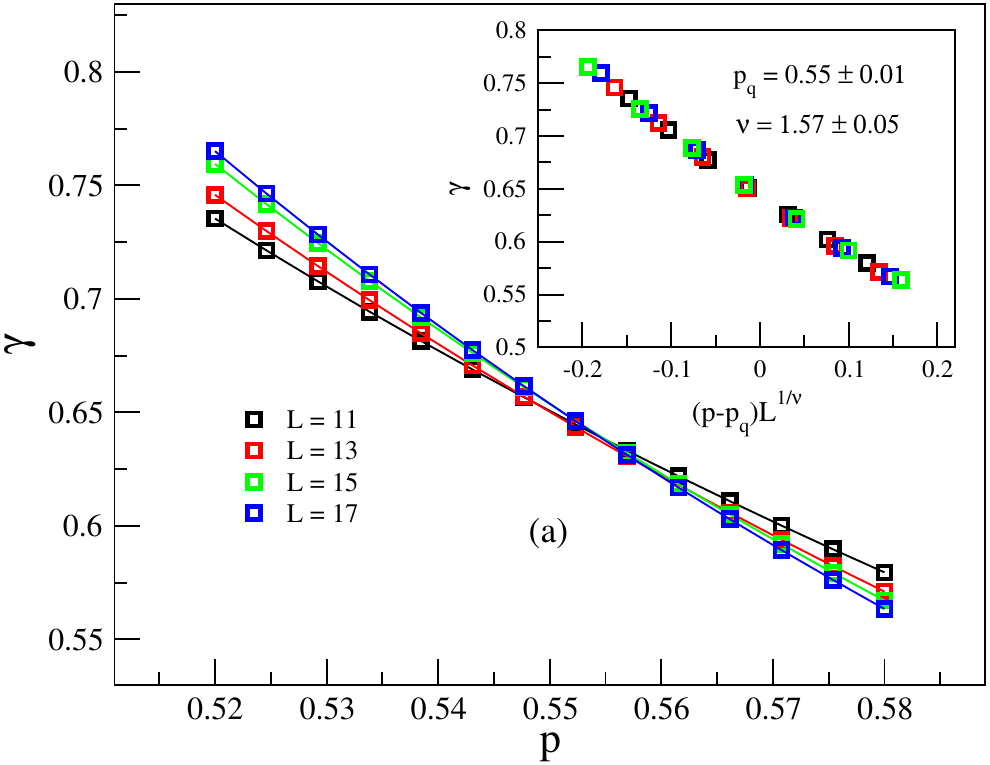}
\includegraphics[scale=0.44]{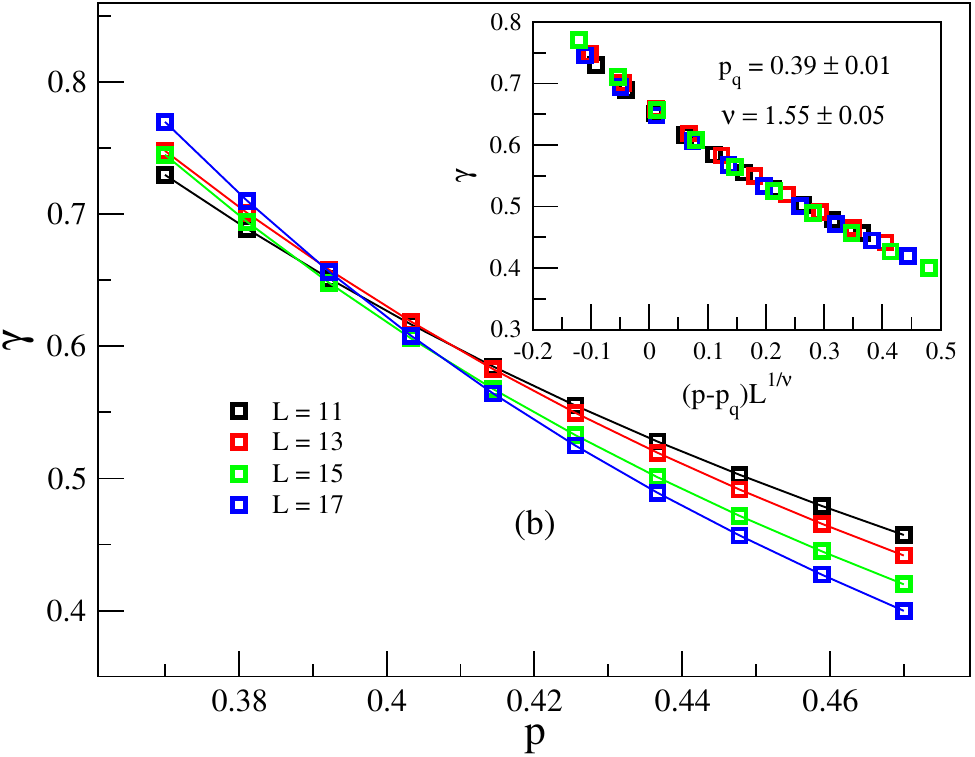}
\caption{Same as Fig.\,\ref{fig:gamma_lieb2D} but for the layered Lieb lattice.}
\label{fig:gamma_layered}
\end{figure}

\begin{figure}[t]
\centering
\includegraphics[scale=0.44]{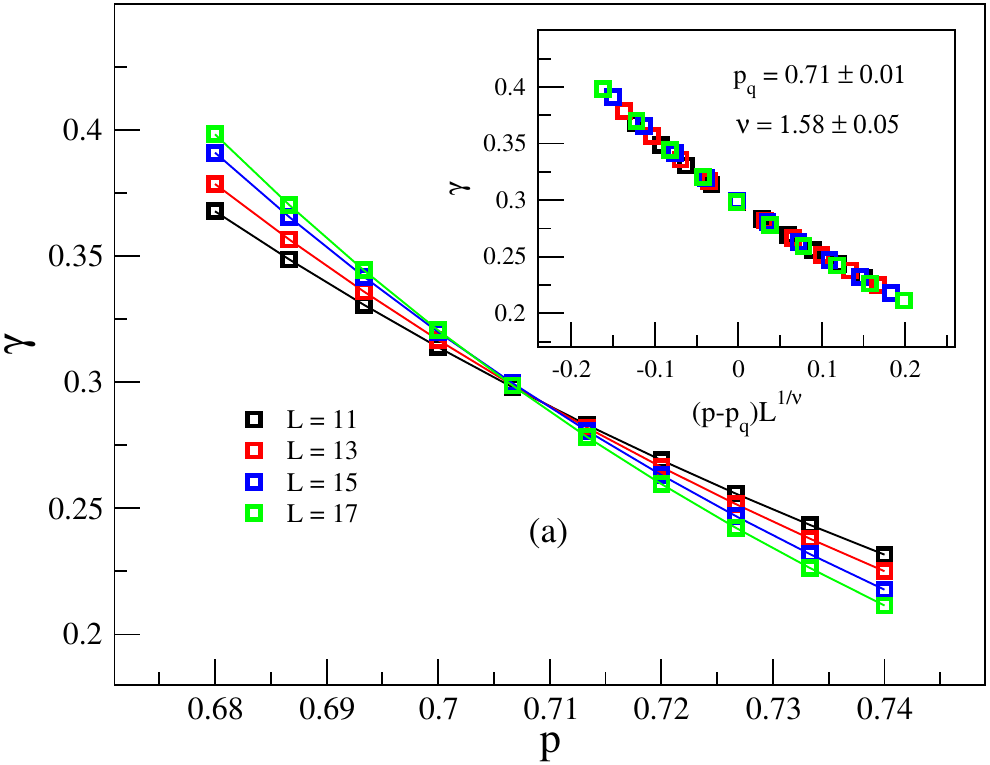}
\includegraphics[scale=0.44]{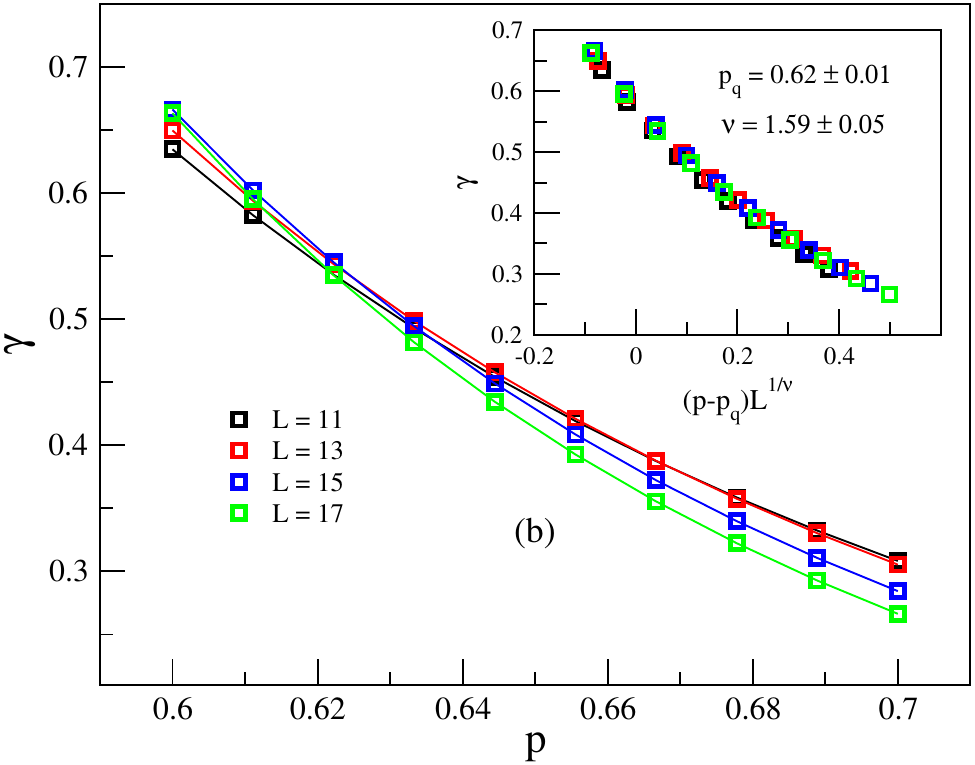}
\caption{Same as Fig.\,\ref{fig:gamma_lieb2D} but for the perovskite lattice.}
\label{fig:Perovskite}
\end{figure}

\subsection{The 3D Lieb lattices}
\label{ssec:3D}

We now discuss the results for the three-dimensional Lieb lattices. 
Their geometry may be described as follows (see Fig.\,\ref{fig:lattices}): 
For the LLL we take $2L$ sites along the $\hat{\mathbf{x}}$ and $\hat{\mathbf{y}}$ directions, and $L$ layers along the $\hat{\mathbf{z}}$ direction. 
For the PL we take $2L$ sites along each of the cartesian directions. 
In our simulations, we consider $M$ disorder realizations, with $M = 900\,(300)$ for the smallest (largest) lattice size, $L = 11\,(17)$.

Figure \ref{fig:gamma_layered} shows plots of $\gamma(p,L)$ as a function of $p$ for the LLL. 
As before, we optimize the data collapse (not shown) through Eqs.\,\eqref{eq. correlationlenght} and \eqref{Eq.gamma_scaling}  to obtain $p_{q}^{\text{(s)}} = 0.55 \pm 0.01$ with $\nu = 1.57 \pm 0.05$, and $p_{q}^{\text{(b)}} = 0.42 \pm 0.01$ with $\nu = 1.55 \pm 0.05$.
Again, $p_q^{\text{(s)}}>p_q^{\text{(b)}}$, and there is no change in universality class.

Let us now present the results for the PL. 
Figure \ref{fig:Perovskite} shows plots of $\gamma(p,L)$ as a function of $p$ for different linear sizes, $L$. 
As before, we optimize the data collapse (not shown) through Eqs.\,\eqref{eq. correlationlenght} and \eqref{Eq.gamma_scaling}  to obtain $p_{q}^{\text{(s)}} = 0.71 \pm 0.01$ with $\nu = 1.58 \pm 0.05$, and $p_{q}^{\text{(b)}} = 0.62 \pm 0.01$ with $\nu = 1.59 \pm 0.05$.
Again, $p_q^{\text{(s)}}>p_q^{\text{(b)}}$, and there is no change in universality class.

\section{Conclusions}
\label{sec:conc}

We have studied the quantum percolation problem on Lieb lattices in two and three dimensions. 
Through random matrix theory we have determined the level spacing distributions, which, aided by finite-size scaling theory, allowed us to obtain accurate estimates for site- and bond percolation thresholds and critical exponents.

From these estimates, we conclude that the thresholds may be ordered as follows:
\begin{equation}
	p_q^\text{SC} < p_q^\text{LLL} < p_q^\text{PL} < p_q^\text{SQ} < p_q^\text{Lieb}, 
\end{equation}
which confirms that even in the quantum case, the threshold is inversely proportional to an average coordination number, as in the classical case \cite{Oliveira21}.

The accuracy in the localization length exponent we obtained allows us to state that: 
(i) for the square and Lieb lattices, the diffusive regime resists to disorder up to a finite threshold, $q_q=1-p_q$;
(ii) bond and site QP problems on Lieb lattices belong to the same universality class;
(iii) $\nu$ decreases with lattice dimensionality, $d$, on Lieb lattices, similarly to the classical percolation problem; 
(iv) for a given $d$, the critical exponent is the same as for the corresponding hypercubic lattice;
(v) $\nu$ for the QP problem is always larger than the one for the corresponding classical percolation problem;
(vi) in three dimensions, QP on Lieb lattices belongs to the same universality class as the AIM.
The lack of emergent new universality classes on Lieb lattices may be attributed to the fact that off-diagonal disorder destroys chiral symmetry \cite{Ramachandran17}.

\section*{ACKNOWLEDGMENTS}

The authors are grateful to N.C.\,Costa for discussions, and to the Brazilian Agencies Conselho Nacional de Desenvolvimento Cient\'\i fico e Tecnol\'ogico (CNPq), and Coordena\c c\~ao de Aperfei\c coamento de Pessoal de Ensino Superior (CAPES) for funding this project. 
R.R.d.S.\,gratefully acknowledges support from Funda\c c\~ao Carlos Chagas de Apoio \`a Pesquisa (FAPERJ), through the grant E-26/210.974/2024, and from CNPq, Grant No.~314611/2023-1.



\bibliography{Qperc.bib}

\begin{thebibliography}{60}%
\makeatletter
\providecommand \@ifxundefined [1]{%
 \@ifx{#1\undefined}
}%
\providecommand \@ifnum [1]{%
 \ifnum #1\expandafter \@firstoftwo
 \else \expandafter \@secondoftwo
 \fi
}%
\providecommand \@ifx [1]{%
 \ifx #1\expandafter \@firstoftwo
 \else \expandafter \@secondoftwo
 \fi
}%
\providecommand \natexlab [1]{#1}%
\providecommand \enquote  [1]{``#1''}%
\providecommand \bibnamefont  [1]{#1}%
\providecommand \bibfnamefont [1]{#1}%
\providecommand \citenamefont [1]{#1}%
\providecommand \href@noop [0]{\@secondoftwo}%
\providecommand \href [0]{\begingroup \@sanitize@url \@href}%
\providecommand \@href[1]{\@@startlink{#1}\@@href}%
\providecommand \@@href[1]{\endgroup#1\@@endlink}%
\providecommand \@sanitize@url [0]{\catcode `\\12\catcode `\$12\catcode `\&12\catcode `\#12\catcode `\^12\catcode `\_12\catcode `\%12\relax}%
\providecommand \@@startlink[1]{}%
\providecommand \@@endlink[0]{}%
\providecommand \url  [0]{\begingroup\@sanitize@url \@url }%
\providecommand \@url [1]{\endgroup\@href {#1}{\urlprefix }}%
\providecommand \urlprefix  [0]{URL }%
\providecommand \Eprint [0]{\href }%
\providecommand \doibase [0]{http://dx.doi.org/}%
\providecommand \selectlanguage [0]{\@gobble}%
\providecommand \bibinfo  [0]{\@secondoftwo}%
\providecommand \bibfield  [0]{\@secondoftwo}%
\providecommand \translation [1]{[#1]}%
\providecommand \BibitemOpen [0]{}%
\providecommand \bibitemStop [0]{}%
\providecommand \bibitemNoStop [0]{.\EOS\space}%
\providecommand \EOS [0]{\spacefactor3000\relax}%
\providecommand \BibitemShut  [1]{\csname bibitem#1\endcsname}%
\let\auto@bib@innerbib\@empty
\bibitem [{\citenamefont {Anderson}(1958)}]{Anderson58}%
  \BibitemOpen
  \bibfield  {author} {\bibinfo {author} {\bibfnamefont {P.~W.}\ \bibnamefont {Anderson}},\ }\bibfield  {title} {\enquote {\bibinfo {title} {Absence of diffusion in certain random lattices},}\ }\href {\doibase 10.1103/PhysRev.109.1492} {\bibfield  {journal} {\bibinfo  {journal} {Phys. Rev.}\ }\textbf {\bibinfo {volume} {109}},\ \bibinfo {pages} {1492--1505} (\bibinfo {year} {1958})}\BibitemShut {NoStop}%
\bibitem [{\citenamefont {Lagendijk}\ \emph {et~al.}(2009)\citenamefont {Lagendijk}, \citenamefont {Tiggelen},\ and\ \citenamefont {Wiersma}}]{Lagendijk09}%
  \BibitemOpen
  \bibfield  {author} {\bibinfo {author} {\bibfnamefont {Ad}~\bibnamefont {Lagendijk}}, \bibinfo {author} {\bibfnamefont {Bart~van}\ \bibnamefont {Tiggelen}}, \ and\ \bibinfo {author} {\bibfnamefont {Diederik~S.}\ \bibnamefont {Wiersma}},\ }\bibfield  {title} {\enquote {\bibinfo {title} {{Fifty years of Anderson localization}},}\ }\href {\doibase 10.1063/1.3206091} {\bibfield  {journal} {\bibinfo  {journal} {Physics Today}\ }\textbf {\bibinfo {volume} {62}},\ \bibinfo {pages} {24--29} (\bibinfo {year} {2009})}\BibitemShut {NoStop}%
\bibitem [{\citenamefont {Abrahams}\ \emph {et~al.}(1979)\citenamefont {Abrahams}, \citenamefont {Anderson}, \citenamefont {Licciardello},\ and\ \citenamefont {Ramakrishnan}}]{Abrahams79}%
  \BibitemOpen
  \bibfield  {author} {\bibinfo {author} {\bibfnamefont {E.}~\bibnamefont {Abrahams}}, \bibinfo {author} {\bibfnamefont {P.~W.}\ \bibnamefont {Anderson}}, \bibinfo {author} {\bibfnamefont {D.~C.}\ \bibnamefont {Licciardello}}, \ and\ \bibinfo {author} {\bibfnamefont {T.~V.}\ \bibnamefont {Ramakrishnan}},\ }\bibfield  {title} {\enquote {\bibinfo {title} {Scaling theory of localization: Absence of quantum diffusion in two dimensions},}\ }\href {\doibase 10.1103/PhysRevLett.42.673} {\bibfield  {journal} {\bibinfo  {journal} {Phys. Rev. Lett.}\ }\textbf {\bibinfo {volume} {42}},\ \bibinfo {pages} {673--676} (\bibinfo {year} {1979})}\BibitemShut {NoStop}%
\bibitem [{\citenamefont {Stauffer}\ and\ \citenamefont {Aharony}(1994)}]{Stauffer94}%
  \BibitemOpen
  \bibfield  {author} {\bibinfo {author} {\bibfnamefont {D.}~\bibnamefont {Stauffer}}\ and\ \bibinfo {author} {\bibfnamefont {A.}~\bibnamefont {Aharony}},\ }\href {https://books.google.com.br/books?id=v66plleij5QC} {\emph {\bibinfo {title} {Introduction To Percolation Theory}}}\ (\bibinfo  {publisher} {Taylor \& Francis},\ \bibinfo {year} {1994})\BibitemShut {NoStop}%
\bibitem [{\citenamefont {Shapir}\ \emph {et~al.}(1982)\citenamefont {Shapir}, \citenamefont {Aharony},\ and\ \citenamefont {Harris}}]{Shapir82}%
  \BibitemOpen
  \bibfield  {author} {\bibinfo {author} {\bibfnamefont {Yonathan}\ \bibnamefont {Shapir}}, \bibinfo {author} {\bibfnamefont {Amnon}\ \bibnamefont {Aharony}}, \ and\ \bibinfo {author} {\bibfnamefont {A.~Brooks}\ \bibnamefont {Harris}},\ }\bibfield  {title} {\enquote {\bibinfo {title} {Localization and quantum percolation},}\ }\href {\doibase 10.1103/PhysRevLett.49.486} {\bibfield  {journal} {\bibinfo  {journal} {Phys. Rev. Lett.}\ }\textbf {\bibinfo {volume} {49}},\ \bibinfo {pages} {486--489} (\bibinfo {year} {1982})}\BibitemShut {NoStop}%
\bibitem [{\citenamefont {Soukoulis}\ and\ \citenamefont {Grest}(1991)}]{Soukoulis91}%
  \BibitemOpen
  \bibfield  {author} {\bibinfo {author} {\bibfnamefont {C.~M.}\ \bibnamefont {Soukoulis}}\ and\ \bibinfo {author} {\bibfnamefont {Gary~S.}\ \bibnamefont {Grest}},\ }\bibfield  {title} {\enquote {\bibinfo {title} {Localization in two-dimensional quantum percolation},}\ }\href {\doibase 10.1103/PhysRevB.44.4685} {\bibfield  {journal} {\bibinfo  {journal} {Phys. Rev. B}\ }\textbf {\bibinfo {volume} {44}},\ \bibinfo {pages} {4685--4688} (\bibinfo {year} {1991})}\BibitemShut {NoStop}%
\bibitem [{\citenamefont {Berkovits}\ and\ \citenamefont {Avishai}(1996)}]{Berkovits96}%
  \BibitemOpen
  \bibfield  {author} {\bibinfo {author} {\bibfnamefont {Richard}\ \bibnamefont {Berkovits}}\ and\ \bibinfo {author} {\bibfnamefont {Yshai}\ \bibnamefont {Avishai}},\ }\bibfield  {title} {\enquote {\bibinfo {title} {Spectral statistics near the quantum percolation threshold},}\ }\href {\doibase 10.1103/PhysRevB.53.R16125} {\bibfield  {journal} {\bibinfo  {journal} {Phys. Rev. B}\ }\textbf {\bibinfo {volume} {53}},\ \bibinfo {pages} {R16125--R16128} (\bibinfo {year} {1996})}\BibitemShut {NoStop}%
\bibitem [{\citenamefont {Nazareno}\ \emph {et~al.}(2002)\citenamefont {Nazareno}, \citenamefont {de~Brito},\ and\ \citenamefont {Rodrigues}}]{Nazareno02}%
  \BibitemOpen
  \bibfield  {author} {\bibinfo {author} {\bibfnamefont {H.~N.}\ \bibnamefont {Nazareno}}, \bibinfo {author} {\bibfnamefont {P.~E.}\ \bibnamefont {de~Brito}}, \ and\ \bibinfo {author} {\bibfnamefont {E.~S.}\ \bibnamefont {Rodrigues}},\ }\bibfield  {title} {\enquote {\bibinfo {title} {Quantum percolation in a two-dimensional finite binary alloy: Interplay between the strength of disorder and alloy composition},}\ }\href {\doibase 10.1103/PhysRevB.66.012205} {\bibfield  {journal} {\bibinfo  {journal} {Phys. Rev. B}\ }\textbf {\bibinfo {volume} {66}},\ \bibinfo {pages} {012205} (\bibinfo {year} {2002})}\BibitemShut {NoStop}%
\bibitem [{\citenamefont {Islam}\ and\ \citenamefont {Nakanishi}(2008)}]{Islam08}%
  \BibitemOpen
  \bibfield  {author} {\bibinfo {author} {\bibfnamefont {M~Fhokrul}\ \bibnamefont {Islam}}\ and\ \bibinfo {author} {\bibfnamefont {Hisao}\ \bibnamefont {Nakanishi}},\ }\bibfield  {title} {\enquote {\bibinfo {title} {Localization-delocalization transition in a two-dimensional quantum percolation model},}\ }\href {\doibase 10.1103/PhysRevE.77.061109} {\bibfield  {journal} {\bibinfo  {journal} {Phys. Rev. E}\ }\textbf {\bibinfo {volume} {77}},\ \bibinfo {pages} {061109} (\bibinfo {year} {2008})}\BibitemShut {NoStop}%
\bibitem [{\citenamefont {Schubert}\ and\ \citenamefont {Fehske}(2008)}]{Schubert08}%
  \BibitemOpen
  \bibfield  {author} {\bibinfo {author} {\bibfnamefont {Gerald}\ \bibnamefont {Schubert}}\ and\ \bibinfo {author} {\bibfnamefont {Holger}\ \bibnamefont {Fehske}},\ }\bibfield  {title} {\enquote {\bibinfo {title} {Dynamical aspects of two-dimensional quantum percolation},}\ }\href {\doibase 10.1103/PhysRevB.77.245130} {\bibfield  {journal} {\bibinfo  {journal} {Phys. Rev. B}\ }\textbf {\bibinfo {volume} {77}},\ \bibinfo {pages} {245130} (\bibinfo {year} {2008})}\BibitemShut {NoStop}%
\bibitem [{\citenamefont {Gong}\ and\ \citenamefont {Tong}(2009)}]{Gong09}%
  \BibitemOpen
  \bibfield  {author} {\bibinfo {author} {\bibfnamefont {Longyan}\ \bibnamefont {Gong}}\ and\ \bibinfo {author} {\bibfnamefont {Peiqing}\ \bibnamefont {Tong}},\ }\bibfield  {title} {\enquote {\bibinfo {title} {Localization-delocalization transitions in a two-dimensional quantum percolation model: von {N}eumann entropy studies},}\ }\href {\doibase 10.1103/PhysRevB.80.174205} {\bibfield  {journal} {\bibinfo  {journal} {Phys. Rev. B}\ }\textbf {\bibinfo {volume} {80}},\ \bibinfo {pages} {174205} (\bibinfo {year} {2009})}\BibitemShut {NoStop}%
\bibitem [{\citenamefont {Dillon}\ and\ \citenamefont {Nakanishi}(2014)}]{Dillon14}%
  \BibitemOpen
  \bibfield  {author} {\bibinfo {author} {\bibfnamefont {Brianna~S}\ \bibnamefont {Dillon}}\ and\ \bibinfo {author} {\bibfnamefont {Hisao}\ \bibnamefont {Nakanishi}},\ }\bibfield  {title} {\enquote {\bibinfo {title} {Localization phase diagram of two-dimensional quantum percolation},}\ }\href {\doibase 10.1140/epjb/e2014-50397-4} {\bibfield  {journal} {\bibinfo  {journal} {The European Physical Journal B}\ }\textbf {\bibinfo {volume} {87}},\ \bibinfo {pages} {1--9} (\bibinfo {year} {2014})}\BibitemShut {NoStop}%
\bibitem [{\citenamefont {{Thouless}}(1972)}]{Thouless72}%
  \BibitemOpen
  \bibfield  {author} {\bibinfo {author} {\bibfnamefont {D.~J.}\ \bibnamefont {{Thouless}}},\ }\bibfield  {title} {\enquote {\bibinfo {title} {{A relation between the density of states and range of localization for one dimensional random systems}},}\ }\href {\doibase 10.1088/0022-3719/5/1/010} {\bibfield  {journal} {\bibinfo  {journal} {Journal of Physics C Solid State Physics}\ }\textbf {\bibinfo {volume} {5}},\ \bibinfo {pages} {77--81} (\bibinfo {year} {1972})}\BibitemShut {NoStop}%
\bibitem [{\citenamefont {Mano}\ and\ \citenamefont {Ohtsuki}(2017)}]{Mano2017}%
  \BibitemOpen
  \bibfield  {author} {\bibinfo {author} {\bibfnamefont {Tomohiro}\ \bibnamefont {Mano}}\ and\ \bibinfo {author} {\bibfnamefont {Tomi}\ \bibnamefont {Ohtsuki}},\ }\bibfield  {title} {\enquote {\bibinfo {title} {Phase diagrams of three-dimensional {A}nderson and quantum percolation models using deep three-dimensional convolutional neural network},}\ }\href {\doibase 10.7566/JPSJ.86.113704} {\bibfield  {journal} {\bibinfo  {journal} {Journal of the Physical Society of Japan}\ }\textbf {\bibinfo {volume} {86}},\ \bibinfo {pages} {113704} (\bibinfo {year} {2017})},\ \Eprint {http://arxiv.org/abs/https://doi.org/10.7566/JPSJ.86.113704} {https://doi.org/10.7566/JPSJ.86.113704} \BibitemShut {NoStop}%
\bibitem [{\citenamefont {Lieb}(1989)}]{Lieb89}%
  \BibitemOpen
  \bibfield  {author} {\bibinfo {author} {\bibfnamefont {Elliott~H.}\ \bibnamefont {Lieb}},\ }\bibfield  {title} {\enquote {\bibinfo {title} {Two theorems on the hubbard model},}\ }\href {\doibase 10.1103/PhysRevLett.62.1201} {\bibfield  {journal} {\bibinfo  {journal} {Phys. Rev. Lett.}\ }\textbf {\bibinfo {volume} {62}},\ \bibinfo {pages} {1201--1204} (\bibinfo {year} {1989})}\BibitemShut {NoStop}%
\bibitem [{\citenamefont {Costa}\ \emph {et~al.}(2016)\citenamefont {Costa}, \citenamefont {Mendes-Santos}, \citenamefont {Paiva}, \citenamefont {Santos},\ and\ \citenamefont {Scalettar}}]{Costa16}%
  \BibitemOpen
  \bibfield  {author} {\bibinfo {author} {\bibfnamefont {Natanael~C.}\ \bibnamefont {Costa}}, \bibinfo {author} {\bibfnamefont {Tiago}\ \bibnamefont {Mendes-Santos}}, \bibinfo {author} {\bibfnamefont {Thereza}\ \bibnamefont {Paiva}}, \bibinfo {author} {\bibfnamefont {Raimundo R.~dos}\ \bibnamefont {Santos}}, \ and\ \bibinfo {author} {\bibfnamefont {Richard~T.}\ \bibnamefont {Scalettar}},\ }\bibfield  {title} {\enquote {\bibinfo {title} {Ferromagnetism beyond {L}ieb's theorem},}\ }\href {\doibase 10.1103/PhysRevB.94.155107} {\bibfield  {journal} {\bibinfo  {journal} {Phys. Rev. B}\ }\textbf {\bibinfo {volume} {94}},\ \bibinfo {pages} {155107} (\bibinfo {year} {2016})}\BibitemShut {NoStop}%
\bibitem [{\citenamefont {Costa}\ \emph {et~al.}(2018)\citenamefont {Costa}, \citenamefont {Ara\'ujo}, \citenamefont {Lima}, \citenamefont {Paiva}, \citenamefont {dos Santos},\ and\ \citenamefont {Scalettar}}]{Costa18}%
  \BibitemOpen
  \bibfield  {author} {\bibinfo {author} {\bibfnamefont {N.~C.}\ \bibnamefont {Costa}}, \bibinfo {author} {\bibfnamefont {M.~V.}\ \bibnamefont {Ara\'ujo}}, \bibinfo {author} {\bibfnamefont {J.~P.}\ \bibnamefont {Lima}}, \bibinfo {author} {\bibfnamefont {T.}~\bibnamefont {Paiva}}, \bibinfo {author} {\bibfnamefont {R.~R.}\ \bibnamefont {dos Santos}}, \ and\ \bibinfo {author} {\bibfnamefont {R.~T.}\ \bibnamefont {Scalettar}},\ }\bibfield  {title} {\enquote {\bibinfo {title} {Compressible ferrimagnetism in the depleted periodic {A}nderson model},}\ }\href {\doibase 10.1103/PhysRevB.97.085123} {\bibfield  {journal} {\bibinfo  {journal} {Phys. Rev. B}\ }\textbf {\bibinfo {volume} {97}},\ \bibinfo {pages} {085123} (\bibinfo {year} {2018})}\BibitemShut {NoStop}%
\bibitem [{\citenamefont {Yarmohammadi}\ and\ \citenamefont {Hoi}(2018)}]{Yarmohammadi18}%
  \BibitemOpen
  \bibfield  {author} {\bibinfo {author} {\bibfnamefont {Mohsen}\ \bibnamefont {Yarmohammadi}}\ and\ \bibinfo {author} {\bibfnamefont {Bui~Dinh}\ \bibnamefont {Hoi}},\ }\bibfield  {title} {\enquote {\bibinfo {title} {A controllable magneto-topological property and band gap engineering in 2d ferromagnetic lieb lattice},}\ }\href {\doibase https://doi.org/10.1016/j.jmmm.2018.05.046} {\bibfield  {journal} {\bibinfo  {journal} {Journal of Magnetism and Magnetic Materials}\ }\textbf {\bibinfo {volume} {464}},\ \bibinfo {pages} {103--107} (\bibinfo {year} {2018})}\BibitemShut {NoStop}%
\bibitem [{\citenamefont {Bouzerar}(2023)}]{Bouzerar23}%
  \BibitemOpen
  \bibfield  {author} {\bibinfo {author} {\bibfnamefont {G.}~\bibnamefont {Bouzerar}},\ }\bibfield  {title} {\enquote {\bibinfo {title} {Flat band induced room-temperature ferromagnetism in two-dimensional systems},}\ }\href {\doibase 10.1103/PhysRevB.107.184441} {\bibfield  {journal} {\bibinfo  {journal} {Phys. Rev. B}\ }\textbf {\bibinfo {volume} {107}},\ \bibinfo {pages} {184441} (\bibinfo {year} {2023})}\BibitemShut {NoStop}%
\bibitem [{\citenamefont {Weeks}\ and\ \citenamefont {Franz}(2010)}]{Weeks10}%
  \BibitemOpen
  \bibfield  {author} {\bibinfo {author} {\bibfnamefont {C.}~\bibnamefont {Weeks}}\ and\ \bibinfo {author} {\bibfnamefont {M.}~\bibnamefont {Franz}},\ }\bibfield  {title} {\enquote {\bibinfo {title} {Topological insulators on the lieb and perovskite lattices},}\ }\href {\doibase 10.1103/PhysRevB.82.085310} {\bibfield  {journal} {\bibinfo  {journal} {Phys. Rev. B}\ }\textbf {\bibinfo {volume} {82}},\ \bibinfo {pages} {085310} (\bibinfo {year} {2010})}\BibitemShut {NoStop}%
\bibitem [{\citenamefont {Jiang}\ \emph {et~al.}(2019)\citenamefont {Jiang}, \citenamefont {Kang}, \citenamefont {Huang}, \citenamefont {Xu}, \citenamefont {Low},\ and\ \citenamefont {Liu}}]{Jiang19}%
  \BibitemOpen
  \bibfield  {author} {\bibinfo {author} {\bibfnamefont {Wei}\ \bibnamefont {Jiang}}, \bibinfo {author} {\bibfnamefont {Meng}\ \bibnamefont {Kang}}, \bibinfo {author} {\bibfnamefont {Huaqing}\ \bibnamefont {Huang}}, \bibinfo {author} {\bibfnamefont {Hongxing}\ \bibnamefont {Xu}}, \bibinfo {author} {\bibfnamefont {Tony}\ \bibnamefont {Low}}, \ and\ \bibinfo {author} {\bibfnamefont {Feng}\ \bibnamefont {Liu}},\ }\bibfield  {title} {\enquote {\bibinfo {title} {Topological band evolution between lieb and kagome lattices},}\ }\href {\doibase 10.1103/PhysRevB.99.125131} {\bibfield  {journal} {\bibinfo  {journal} {Phys. Rev. B}\ }\textbf {\bibinfo {volume} {99}},\ \bibinfo {pages} {125131} (\bibinfo {year} {2019})}\BibitemShut {NoStop}%
\bibitem [{\citenamefont {Jiang}\ \emph {et~al.}(2020)\citenamefont {Jiang}, \citenamefont {Zhang}, \citenamefont {Wang}, \citenamefont {Liu},\ and\ \citenamefont {Low}}]{Jiang20}%
  \BibitemOpen
  \bibfield  {author} {\bibinfo {author} {\bibfnamefont {Wei}\ \bibnamefont {Jiang}}, \bibinfo {author} {\bibfnamefont {Shunhong}\ \bibnamefont {Zhang}}, \bibinfo {author} {\bibfnamefont {Zhengfei}\ \bibnamefont {Wang}}, \bibinfo {author} {\bibfnamefont {Feng}\ \bibnamefont {Liu}}, \ and\ \bibinfo {author} {\bibfnamefont {Tony}\ \bibnamefont {Low}},\ }\bibfield  {title} {\enquote {\bibinfo {title} {Topological band engineering of lieb lattice in phthalocyanine-based metal–organic frameworks},}\ }\href {\doibase 10.1021/acs.nanolett.9b05242} {\bibfield  {journal} {\bibinfo  {journal} {Nano Letters}\ }\textbf {\bibinfo {volume} {20}},\ \bibinfo {pages} {1959--1966} (\bibinfo {year} {2020})}\BibitemShut {NoStop}%
\bibitem [{\citenamefont {Poata}\ \emph {et~al.}(2023)\citenamefont {Poata}, \citenamefont {Taddei},\ and\ \citenamefont {Governale}}]{Poata23}%
  \BibitemOpen
  \bibfield  {author} {\bibinfo {author} {\bibfnamefont {Joseph}\ \bibnamefont {Poata}}, \bibinfo {author} {\bibfnamefont {Fabio}\ \bibnamefont {Taddei}}, \ and\ \bibinfo {author} {\bibfnamefont {Michele}\ \bibnamefont {Governale}},\ }\bibfield  {title} {\enquote {\bibinfo {title} {Corner states of two-dimensional second-order topological insulators with a chiral symmetry and broken time reversal and charge conjugation},}\ }\href {\doibase 10.1103/PhysRevB.108.115405} {\bibfield  {journal} {\bibinfo  {journal} {Phys. Rev. B}\ }\textbf {\bibinfo {volume} {108}},\ \bibinfo {pages} {115405} (\bibinfo {year} {2023})}\BibitemShut {NoStop}%
\bibitem [{\citenamefont {Andrade}\ \emph {et~al.}(2023)\citenamefont {Andrade}, \citenamefont {L\'opez-Ur\'{\i}as},\ and\ \citenamefont {Naumis}}]{Andrade23}%
  \BibitemOpen
  \bibfield  {author} {\bibinfo {author} {\bibfnamefont {Elias}\ \bibnamefont {Andrade}}, \bibinfo {author} {\bibfnamefont {Florentino}\ \bibnamefont {L\'opez-Ur\'{\i}as}}, \ and\ \bibinfo {author} {\bibfnamefont {Gerardo~G.}\ \bibnamefont {Naumis}},\ }\bibfield  {title} {\enquote {\bibinfo {title} {Topological origin of flat bands as pseudo-landau levels in uniaxial strained graphene nanoribbons and induced magnetic ordering due to electron-electron interactions},}\ }\href {\doibase 10.1103/PhysRevB.107.235143} {\bibfield  {journal} {\bibinfo  {journal} {Phys. Rev. B}\ }\textbf {\bibinfo {volume} {107}},\ \bibinfo {pages} {235143} (\bibinfo {year} {2023})}\BibitemShut {NoStop}%
\bibitem [{\citenamefont {Noda}\ \emph {et~al.}(2015)\citenamefont {Noda}, \citenamefont {Inaba},\ and\ \citenamefont {Yamashita}}]{Noda15}%
  \BibitemOpen
  \bibfield  {author} {\bibinfo {author} {\bibfnamefont {Kazuto}\ \bibnamefont {Noda}}, \bibinfo {author} {\bibfnamefont {Kensuke}\ \bibnamefont {Inaba}}, \ and\ \bibinfo {author} {\bibfnamefont {Makoto}\ \bibnamefont {Yamashita}},\ }\bibfield  {title} {\enquote {\bibinfo {title} {Magnetism in the three-dimensional layered {L}ieb lattice: Enhanced transition temperature via flat-band and van {H}ove singularities},}\ }\href {\doibase 10.1103/PhysRevA.91.063610} {\bibfield  {journal} {\bibinfo  {journal} {Phys. Rev. A}\ }\textbf {\bibinfo {volume} {91}},\ \bibinfo {pages} {063610} (\bibinfo {year} {2015})}\BibitemShut {NoStop}%
\bibitem [{\citenamefont {Liu}\ \emph {et~al.}(2021)\citenamefont {Liu}, \citenamefont {Mao}, \citenamefont {Zhong},\ and\ \citenamefont {Römer}}]{Liu21}%
  \BibitemOpen
  \bibfield  {author} {\bibinfo {author} {\bibfnamefont {Jie}\ \bibnamefont {Liu}}, \bibinfo {author} {\bibfnamefont {Xiaoyu}\ \bibnamefont {Mao}}, \bibinfo {author} {\bibfnamefont {Jianxin}\ \bibnamefont {Zhong}}, \ and\ \bibinfo {author} {\bibfnamefont {Rudolf~A.}\ \bibnamefont {Römer}},\ }\bibfield  {title} {\enquote {\bibinfo {title} {Localization properties in lieb lattices and their extensions},}\ }\href {\doibase https://doi.org/10.1016/j.aop.2021.168544} {\bibfield  {journal} {\bibinfo  {journal} {Annals of Physics}\ }\textbf {\bibinfo {volume} {435}},\ \bibinfo {pages} {168544} (\bibinfo {year} {2021})},\ \bibinfo {note} {special Issue on Localisation 2020}\BibitemShut {NoStop}%
\bibitem [{\citenamefont {Liu}\ \emph {et~al.}(2020)\citenamefont {Liu}, \citenamefont {Mao}, \citenamefont {Zhong},\ and\ \citenamefont {R\"omer}}]{Liu20}%
  \BibitemOpen
  \bibfield  {author} {\bibinfo {author} {\bibfnamefont {Jie}\ \bibnamefont {Liu}}, \bibinfo {author} {\bibfnamefont {Xiaoyu}\ \bibnamefont {Mao}}, \bibinfo {author} {\bibfnamefont {Jianxin}\ \bibnamefont {Zhong}}, \ and\ \bibinfo {author} {\bibfnamefont {Rudolf~A.}\ \bibnamefont {R\"omer}},\ }\bibfield  {title} {\enquote {\bibinfo {title} {Localization, phases, and transitions in three-dimensional extended lieb lattices},}\ }\href {\doibase 10.1103/PhysRevB.102.174207} {\bibfield  {journal} {\bibinfo  {journal} {Phys. Rev. B}\ }\textbf {\bibinfo {volume} {102}},\ \bibinfo {pages} {174207} (\bibinfo {year} {2020})}\BibitemShut {NoStop}%
\bibitem [{\citenamefont {Oliveira}\ \emph {et~al.}(2021)\citenamefont {Oliveira}, \citenamefont {de~Lima}, \citenamefont {Costa},\ and\ \citenamefont {dos Santos}}]{Oliveira21}%
  \BibitemOpen
  \bibfield  {author} {\bibinfo {author} {\bibfnamefont {W.~S.}\ \bibnamefont {Oliveira}}, \bibinfo {author} {\bibfnamefont {J.~Pimentel}\ \bibnamefont {de~Lima}}, \bibinfo {author} {\bibfnamefont {N.~C.}\ \bibnamefont {Costa}}, \ and\ \bibinfo {author} {\bibfnamefont {R.~R.}\ \bibnamefont {dos Santos}},\ }\bibfield  {title} {\enquote {\bibinfo {title} {Percolation on {L}ieb lattices},}\ }\href {\doibase 10.1103/PhysRevE.104.064122} {\bibfield  {journal} {\bibinfo  {journal} {Phys. Rev. E}\ }\textbf {\bibinfo {volume} {104}},\ \bibinfo {pages} {064122} (\bibinfo {year} {2021})}\BibitemShut {NoStop}%
\bibitem [{\citenamefont {Andreev}\ \emph {et~al.}(1996)\citenamefont {Andreev}, \citenamefont {Agam}, \citenamefont {Simons},\ and\ \citenamefont {Altshuler}}]{Andreev96}%
  \BibitemOpen
  \bibfield  {author} {\bibinfo {author} {\bibfnamefont {A.~V.}\ \bibnamefont {Andreev}}, \bibinfo {author} {\bibfnamefont {O.}~\bibnamefont {Agam}}, \bibinfo {author} {\bibfnamefont {B.~D.}\ \bibnamefont {Simons}}, \ and\ \bibinfo {author} {\bibfnamefont {B.~L.}\ \bibnamefont {Altshuler}},\ }\bibfield  {title} {\enquote {\bibinfo {title} {Quantum chaos, irreversible classical dynamics, and random matrix theory},}\ }\href {\doibase 10.1103/PhysRevLett.76.3947} {\bibfield  {journal} {\bibinfo  {journal} {Phys. Rev. Lett.}\ }\textbf {\bibinfo {volume} {76}},\ \bibinfo {pages} {3947--3950} (\bibinfo {year} {1996})}\BibitemShut {NoStop}%
\bibitem [{\citenamefont {Muttalib}\ \emph {et~al.}(1987)\citenamefont {Muttalib}, \citenamefont {Pichard},\ and\ \citenamefont {Stone}}]{Muttalib87}%
  \BibitemOpen
  \bibfield  {author} {\bibinfo {author} {\bibfnamefont {K.~A.}\ \bibnamefont {Muttalib}}, \bibinfo {author} {\bibfnamefont {J.~L.}\ \bibnamefont {Pichard}}, \ and\ \bibinfo {author} {\bibfnamefont {A.~Douglas}\ \bibnamefont {Stone}},\ }\bibfield  {title} {\enquote {\bibinfo {title} {Random-matrix theory and universal statistics for disordered quantum conductors},}\ }\href {\doibase 10.1103/PhysRevLett.59.2475} {\bibfield  {journal} {\bibinfo  {journal} {Phys. Rev. Lett.}\ }\textbf {\bibinfo {volume} {59}},\ \bibinfo {pages} {2475--2478} (\bibinfo {year} {1987})}\BibitemShut {NoStop}%
\bibitem [{\citenamefont {Beenakker}(1997)}]{Beenakker97}%
  \BibitemOpen
  \bibfield  {author} {\bibinfo {author} {\bibfnamefont {C.~W.~J.}\ \bibnamefont {Beenakker}},\ }\bibfield  {title} {\enquote {\bibinfo {title} {Random-matrix theory of quantum transport},}\ }\href {\doibase 10.1103/RevModPhys.69.731} {\bibfield  {journal} {\bibinfo  {journal} {Rev. Mod. Phys.}\ }\textbf {\bibinfo {volume} {69}},\ \bibinfo {pages} {731--808} (\bibinfo {year} {1997})}\BibitemShut {NoStop}%
\bibitem [{\citenamefont {Wigner}(1957)}]{wigner57}%
  \BibitemOpen
  \bibfield  {author} {\bibinfo {author} {\bibfnamefont {EP}~\bibnamefont {Wigner}},\ }\bibfield  {title} {\enquote {\bibinfo {title} {Can. math. congr. proc.}}\ }\href@noop {} {\bibfield  {journal} {\bibinfo  {journal} {U of Toronto Press, Toronto}\ ,\ \bibinfo {pages} {174}} (\bibinfo {year} {1957})}\BibitemShut {NoStop}%
\bibitem [{\citenamefont {Shklovskii}\ \emph {et~al.}(1993)\citenamefont {Shklovskii}, \citenamefont {Shapiro}, \citenamefont {Sears}, \citenamefont {Lambrianides},\ and\ \citenamefont {Shore}}]{Shklovskii93}%
  \BibitemOpen
  \bibfield  {author} {\bibinfo {author} {\bibfnamefont {B.~I.}\ \bibnamefont {Shklovskii}}, \bibinfo {author} {\bibfnamefont {B.}~\bibnamefont {Shapiro}}, \bibinfo {author} {\bibfnamefont {B.~R.}\ \bibnamefont {Sears}}, \bibinfo {author} {\bibfnamefont {P.}~\bibnamefont {Lambrianides}}, \ and\ \bibinfo {author} {\bibfnamefont {H.~B.}\ \bibnamefont {Shore}},\ }\bibfield  {title} {\enquote {\bibinfo {title} {Statistics of spectra of disordered systems near the metal-insulator transition},}\ }\href {\doibase 10.1103/PhysRevB.47.11487} {\bibfield  {journal} {\bibinfo  {journal} {Phys. Rev. B}\ }\textbf {\bibinfo {volume} {47}},\ \bibinfo {pages} {11487--11490} (\bibinfo {year} {1993})}\BibitemShut {NoStop}%
\bibitem [{\citenamefont {Kottos}\ and\ \citenamefont {Weiss}(2002)}]{Kottos02}%
  \BibitemOpen
  \bibfield  {author} {\bibinfo {author} {\bibfnamefont {Tsampikos}\ \bibnamefont {Kottos}}\ and\ \bibinfo {author} {\bibfnamefont {Matthias}\ \bibnamefont {Weiss}},\ }\bibfield  {title} {\enquote {\bibinfo {title} {Statistics of resonances and delay times: A criterion for metal-insulator transitions},}\ }\href {\doibase 10.1103/PhysRevLett.89.056401} {\bibfield  {journal} {\bibinfo  {journal} {Phys. Rev. Lett.}\ }\textbf {\bibinfo {volume} {89}},\ \bibinfo {pages} {056401} (\bibinfo {year} {2002})}\BibitemShut {NoStop}%
\bibitem [{\citenamefont {Zharekeshev}\ and\ \citenamefont {Kramer}(1995)}]{Zharekeshev95}%
  \BibitemOpen
  \bibfield  {author} {\bibinfo {author} {\bibfnamefont {I.~Kh.}\ \bibnamefont {Zharekeshev}}\ and\ \bibinfo {author} {\bibfnamefont {B.}~\bibnamefont {Kramer}},\ }\bibfield  {title} {\enquote {\bibinfo {title} {Scaling of level statistics at the disorder-induced metal-insulator transition},}\ }\href {\doibase 10.1103/PhysRevB.51.17239} {\bibfield  {journal} {\bibinfo  {journal} {Phys. Rev. B}\ }\textbf {\bibinfo {volume} {51}},\ \bibinfo {pages} {17239--17242} (\bibinfo {year} {1995})}\BibitemShut {NoStop}%
\bibitem [{\citenamefont {Nishigaki}(1999)}]{Nishigaki99}%
  \BibitemOpen
  \bibfield  {author} {\bibinfo {author} {\bibfnamefont {Shinsuke~M.}\ \bibnamefont {Nishigaki}},\ }\bibfield  {title} {\enquote {\bibinfo {title} {Level spacings at the metal-insulator transition in the anderson hamiltonians and multifractal random matrix ensembles},}\ }\href {\doibase 10.1103/PhysRevE.59.2853} {\bibfield  {journal} {\bibinfo  {journal} {Phys. Rev. E}\ }\textbf {\bibinfo {volume} {59}},\ \bibinfo {pages} {2853--2862} (\bibinfo {year} {1999})}\BibitemShut {NoStop}%
\bibitem [{\citenamefont {Travěnec}(2008)}]{Travenec08}%
  \BibitemOpen
  \bibfield  {author} {\bibinfo {author} {\bibfnamefont {Igor}\ \bibnamefont {Travěnec}},\ }\bibfield  {title} {\enquote {\bibinfo {title} {Metal–insulator transition in 3d quantum percolation},}\ }\href@noop {} {\bibfield  {journal} {\bibinfo  {journal} {International Journal of Modern Physics B}\ }\textbf {\bibinfo {volume} {22}},\ \bibinfo {pages} {5217--5227} (\bibinfo {year} {2008})}\BibitemShut {NoStop}%
\bibitem [{\citenamefont {Cao}\ and\ \citenamefont {Schwarz}(2012)}]{Cao12}%
  \BibitemOpen
  \bibfield  {author} {\bibinfo {author} {\bibfnamefont {L.}~\bibnamefont {Cao}}\ and\ \bibinfo {author} {\bibfnamefont {J.~M.}\ \bibnamefont {Schwarz}},\ }\bibfield  {title} {\enquote {\bibinfo {title} {Level statistics for quantum $k$-core percolation},}\ }\href {\doibase 10.1103/PhysRevB.86.064206} {\bibfield  {journal} {\bibinfo  {journal} {Phys. Rev. B}\ }\textbf {\bibinfo {volume} {86}},\ \bibinfo {pages} {064206} (\bibinfo {year} {2012})}\BibitemShut {NoStop}%
\bibitem [{\citenamefont {Hoshen}\ and\ \citenamefont {Kopelman}(1976)}]{Hoshen76}%
  \BibitemOpen
  \bibfield  {author} {\bibinfo {author} {\bibfnamefont {J.}~\bibnamefont {Hoshen}}\ and\ \bibinfo {author} {\bibfnamefont {R.}~\bibnamefont {Kopelman}},\ }\bibfield  {title} {\enquote {\bibinfo {title} {Percolation and cluster distribution. i. cluster multiple labeling technique and critical concentration algorithm},}\ }\href {\doibase 10.1103/PhysRevB.14.3438} {\bibfield  {journal} {\bibinfo  {journal} {Phys. Rev. B}\ }\textbf {\bibinfo {volume} {14}},\ \bibinfo {pages} {3438--3445} (\bibinfo {year} {1976})}\BibitemShut {NoStop}%
\bibitem [{\citenamefont {H{\"a}ggstr{\"o}m}\ and\ \citenamefont {Jonasson}(2006)}]{Haggstrom06}%
  \BibitemOpen
  \bibfield  {author} {\bibinfo {author} {\bibfnamefont {Olle}\ \bibnamefont {H{\"a}ggstr{\"o}m}}\ and\ \bibinfo {author} {\bibfnamefont {Johan}\ \bibnamefont {Jonasson}},\ }\bibfield  {title} {\enquote {\bibinfo {title} {{Uniqueness and non-uniqueness in percolation theory}},}\ }\href {\doibase 10.1214/154957806000000096} {\bibfield  {journal} {\bibinfo  {journal} {Probability Surveys}\ }\textbf {\bibinfo {volume} {3}},\ \bibinfo {pages} {289 -- 344} (\bibinfo {year} {2006})}\BibitemShut {NoStop}%
\bibitem [{\citenamefont {Mirlin}(2000)}]{Mirlin00}%
  \BibitemOpen
  \bibfield  {author} {\bibinfo {author} {\bibfnamefont {Alexander~D.}\ \bibnamefont {Mirlin}},\ }\bibfield  {title} {\enquote {\bibinfo {title} {Statistics of energy levels and eigenfunctions in disordered systems},}\ }\href {\doibase https://doi.org/10.1016/S0370-1573(99)00091-5} {\bibfield  {journal} {\bibinfo  {journal} {Physics Reports}\ }\textbf {\bibinfo {volume} {326}},\ \bibinfo {pages} {259--382} (\bibinfo {year} {2000})}\BibitemShut {NoStop}%
\bibitem [{\citenamefont {{Al'tshuler}}\ and\ \citenamefont {{Shklovskii}}(1986)}]{Altshuler86}%
  \BibitemOpen
  \bibfield  {author} {\bibinfo {author} {\bibfnamefont {O.~L.}\ \bibnamefont {{Al'tshuler}}}\ and\ \bibinfo {author} {\bibfnamefont {B.~I.}\ \bibnamefont {{Shklovskii}}},\ }\bibfield  {title} {\enquote {\bibinfo {title} {{Repulsion of energy levels and conductivity of small metal samples}},}\ }\href@noop {} {\bibfield  {journal} {\bibinfo  {journal} {Soviet Journal of Experimental and Theoretical Physics}\ }\textbf {\bibinfo {volume} {64}},\ \bibinfo {pages} {127} (\bibinfo {year} {1986})}\BibitemShut {NoStop}%
\bibitem [{\citenamefont {Kaneko}\ and\ \citenamefont {Ohtsuki}(1999)}]{Kaneko99}%
  \BibitemOpen
  \bibfield  {author} {\bibinfo {author} {\bibfnamefont {Atsushi}\ \bibnamefont {Kaneko}}\ and\ \bibinfo {author} {\bibfnamefont {Tomi}\ \bibnamefont {Ohtsuki}},\ }\bibfield  {title} {\enquote {\bibinfo {title} {Three-dimensional quantum percolation studied by level statistics},}\ }\href {\doibase 10.1143/JPSJ.68.1488} {\bibfield  {journal} {\bibinfo  {journal} {Journal of the Physical Society of Japan}\ }\textbf {\bibinfo {volume} {68}},\ \bibinfo {pages} {1488--1491} (\bibinfo {year} {1999})}\BibitemShut {NoStop}%
\bibitem [{\citenamefont {Fisher}(1971)}]{Fisher71}%
  \BibitemOpen
  \bibfield  {author} {\bibinfo {author} {\bibfnamefont {M.~E.}\ \bibnamefont {Fisher}},\ }\bibfield  {title} {\enquote {\bibinfo {title} {Critical {P}henomena},}\ }in\ \href@noop {} {\emph {\bibinfo {booktitle} {{P}roceedings of the {E}nrico {F}ermi {I}nternational {S}chool of {P}hysics}}},\ Vol.~\bibinfo {volume} {51},\ \bibinfo {editor} {edited by\ \bibinfo {editor} {\bibfnamefont {M.~S.}\ \bibnamefont {Green}}}\ (\bibinfo  {publisher} {Academic Press, New York},\ \bibinfo {year} {1971})\BibitemShut {NoStop}%
\bibitem [{\citenamefont {Barber}(1983)}]{Barber83}%
  \BibitemOpen
  \bibfield  {author} {\bibinfo {author} {\bibfnamefont {M.~N.}\ \bibnamefont {Barber}},\ }\bibfield  {title} {\enquote {\bibinfo {title} {Finite-size scaling},}\ }in\ \href@noop {} {\emph {\bibinfo {booktitle} {Phase Transitions and Critical Phenomena}}},\ Vol.~\bibinfo {volume} {8},\ \bibinfo {editor} {edited by\ \bibinfo {editor} {\bibfnamefont {Cyril}\ \bibnamefont {Domb}}\ and\ \bibinfo {editor} {\bibfnamefont {Joel~L.}\ \bibnamefont {Lebowitz}}}\ (\bibinfo  {publisher} {Academic Press},\ \bibinfo {address} {New York},\ \bibinfo {year} {1983})\ p.\ \bibinfo {pages} {145}\BibitemShut {NoStop}%
\bibitem [{\citenamefont {dos Santos}\ and\ \citenamefont {Sneddon}(1981)}]{dosSantos81a}%
  \BibitemOpen
  \bibfield  {author} {\bibinfo {author} {\bibfnamefont {R.~R.}\ \bibnamefont {dos Santos}}\ and\ \bibinfo {author} {\bibfnamefont {L.}~\bibnamefont {Sneddon}},\ }\bibfield  {title} {\enquote {\bibinfo {title} {Finite-size rescaling transformations},}\ }\href {\doibase 10.1103/PhysRevB.23.3541} {\bibfield  {journal} {\bibinfo  {journal} {Phys. Rev. B}\ }\textbf {\bibinfo {volume} {23}},\ \bibinfo {pages} {3541--3546} (\bibinfo {year} {1981})}\BibitemShut {NoStop}%
\bibitem [{\citenamefont {Lee}(2008)}]{Lee08}%
  \BibitemOpen
  \bibfield  {author} {\bibinfo {author} {\bibfnamefont {Michael~J.}\ \bibnamefont {Lee}},\ }\bibfield  {title} {\enquote {\bibinfo {title} {Pseudo-random-number generators and the square site percolation threshold},}\ }\href {\doibase 10.1103/PhysRevE.78.031131} {\bibfield  {journal} {\bibinfo  {journal} {Phys. Rev. E}\ }\textbf {\bibinfo {volume} {78}},\ \bibinfo {pages} {031131} (\bibinfo {year} {2008})}\BibitemShut {NoStop}%
\bibitem [{\citenamefont {Stauffer}\ and\ \citenamefont {Aharony}(2018)}]{stauffer18}%
  \BibitemOpen
  \bibfield  {author} {\bibinfo {author} {\bibfnamefont {Dietrich}\ \bibnamefont {Stauffer}}\ and\ \bibinfo {author} {\bibfnamefont {Ammon}\ \bibnamefont {Aharony}},\ }\href@noop {} {\emph {\bibinfo {title} {Introduction to percolation theory}}}\ (\bibinfo  {publisher} {Taylor \& Francis},\ \bibinfo {year} {2018})\BibitemShut {NoStop}%
\bibitem [{\citenamefont {Daboul}\ \emph {et~al.}(2000)\citenamefont {Daboul}, \citenamefont {Chang},\ and\ \citenamefont {Aharony}}]{Daboul00}%
  \BibitemOpen
  \bibfield  {author} {\bibinfo {author} {\bibfnamefont {D.}~\bibnamefont {Daboul}}, \bibinfo {author} {\bibfnamefont {I.}~\bibnamefont {Chang}}, \ and\ \bibinfo {author} {\bibfnamefont {A.}~\bibnamefont {Aharony}},\ }\bibfield  {title} {\enquote {\bibinfo {title} {Series expansion study of quantum percolation on the square lattice},}\ }\href {\doibase https://doi.org/10.1007/PL00011059} {\bibfield  {journal} {\bibinfo  {journal} {European Physical Journal B}\ }\textbf {\bibinfo {volume} {16}},\ \bibinfo {pages} {303--316} (\bibinfo {year} {2000})}\BibitemShut {NoStop}%
\bibitem [{\citenamefont {Wang}\ \emph {et~al.}(2013)\citenamefont {Wang}, \citenamefont {Zhou}, \citenamefont {Zhang}, \citenamefont {Garoni},\ and\ \citenamefont {Deng}}]{Wang13}%
  \BibitemOpen
  \bibfield  {author} {\bibinfo {author} {\bibfnamefont {Junfeng}\ \bibnamefont {Wang}}, \bibinfo {author} {\bibfnamefont {Zongzheng}\ \bibnamefont {Zhou}}, \bibinfo {author} {\bibfnamefont {Wei}\ \bibnamefont {Zhang}}, \bibinfo {author} {\bibfnamefont {Timothy~M.}\ \bibnamefont {Garoni}}, \ and\ \bibinfo {author} {\bibfnamefont {Youjin}\ \bibnamefont {Deng}},\ }\bibfield  {title} {\enquote {\bibinfo {title} {Bond and site percolation in three dimensions},}\ }\href {\doibase 10.1103/PhysRevE.87.052107} {\bibfield  {journal} {\bibinfo  {journal} {Phys. Rev. E}\ }\textbf {\bibinfo {volume} {87}},\ \bibinfo {pages} {052107} (\bibinfo {year} {2013})}\BibitemShut {NoStop}%
\bibitem [{\citenamefont {Soukoulis}\ \emph {et~al.}(1992)\citenamefont {Soukoulis}, \citenamefont {Li},\ and\ \citenamefont {Grest}}]{Soukoulis92}%
  \BibitemOpen
  \bibfield  {author} {\bibinfo {author} {\bibfnamefont {C.~M.}\ \bibnamefont {Soukoulis}}, \bibinfo {author} {\bibfnamefont {Qiming}\ \bibnamefont {Li}}, \ and\ \bibinfo {author} {\bibfnamefont {Gary~S.}\ \bibnamefont {Grest}},\ }\bibfield  {title} {\enquote {\bibinfo {title} {Quantum percolation in three-dimensional systems},}\ }\href {\doibase 10.1103/PhysRevB.45.7724} {\bibfield  {journal} {\bibinfo  {journal} {Phys. Rev. B}\ }\textbf {\bibinfo {volume} {45}},\ \bibinfo {pages} {7724--7729} (\bibinfo {year} {1992})}\BibitemShut {NoStop}%
\bibitem [{\citenamefont {Stadler}\ \emph {et~al.}(1996)\citenamefont {Stadler}, \citenamefont {Kusy},\ and\ \citenamefont {Sikora}}]{Stadler96}%
  \BibitemOpen
  \bibfield  {author} {\bibinfo {author} {\bibfnamefont {A~W}\ \bibnamefont {Stadler}}, \bibinfo {author} {\bibfnamefont {A}~\bibnamefont {Kusy}}, \ and\ \bibinfo {author} {\bibfnamefont {R}~\bibnamefont {Sikora}},\ }\bibfield  {title} {\enquote {\bibinfo {title} {Numerical studies of the anderson transition in three-dimensional quantum site percolation},}\ }\href {\doibase 10.1088/0953-8984/8/17/010} {\bibfield  {journal} {\bibinfo  {journal} {Journal of Physics: Condensed Matter}\ }\textbf {\bibinfo {volume} {8}},\ \bibinfo {pages} {2981} (\bibinfo {year} {1996})}\BibitemShut {NoStop}%
\bibitem [{\citenamefont {Koslowski}\ and\ \citenamefont {von Niessen}(1990)}]{Koslowski90}%
  \BibitemOpen
  \bibfield  {author} {\bibinfo {author} {\bibfnamefont {Th.}\ \bibnamefont {Koslowski}}\ and\ \bibinfo {author} {\bibfnamefont {W.}~\bibnamefont {von Niessen}},\ }\bibfield  {title} {\enquote {\bibinfo {title} {Mobility edges for the quantum percolation problem in two and three dimensions},}\ }\href {\doibase 10.1103/PhysRevB.42.10342} {\bibfield  {journal} {\bibinfo  {journal} {Phys. Rev. B}\ }\textbf {\bibinfo {volume} {42}},\ \bibinfo {pages} {10342--10347} (\bibinfo {year} {1990})}\BibitemShut {NoStop}%
\bibitem [{\citenamefont {Ujfalusi}\ and\ \citenamefont {Varga}(2014)}]{Ujfalusi14}%
  \BibitemOpen
  \bibfield  {author} {\bibinfo {author} {\bibfnamefont {L\'aszl\'o}\ \bibnamefont {Ujfalusi}}\ and\ \bibinfo {author} {\bibfnamefont {Imre}\ \bibnamefont {Varga}},\ }\bibfield  {title} {\enquote {\bibinfo {title} {Quantum percolation transition in three dimensions: Density of states, finite-size scaling, and multifractality},}\ }\href {\doibase 10.1103/PhysRevB.90.174203} {\bibfield  {journal} {\bibinfo  {journal} {Phys. Rev. B}\ }\textbf {\bibinfo {volume} {90}},\ \bibinfo {pages} {174203} (\bibinfo {year} {2014})}\BibitemShut {NoStop}%
\bibitem [{\citenamefont {Meir}\ \emph {et~al.}(1989)\citenamefont {Meir}, \citenamefont {Aharony},\ and\ \citenamefont {Harris}}]{Meir89}%
  \BibitemOpen
  \bibfield  {author} {\bibinfo {author} {\bibfnamefont {Y.}~\bibnamefont {Meir}}, \bibinfo {author} {\bibfnamefont {A.}~\bibnamefont {Aharony}}, \ and\ \bibinfo {author} {\bibfnamefont {A.~Brooks}\ \bibnamefont {Harris}},\ }\bibfield  {title} {\enquote {\bibinfo {title} {Delocalization transition in two-dimensional quantum percolation},}\ }\href {\doibase 10.1209/0295-5075/10/3/015} {\bibfield  {journal} {\bibinfo  {journal} {Europhysics Letters}\ }\textbf {\bibinfo {volume} {10}},\ \bibinfo {pages} {275} (\bibinfo {year} {1989})}\BibitemShut {NoStop}%
\bibitem [{\citenamefont {Odagaki}\ and\ \citenamefont {Chang}(1984)}]{Odagaki84}%
  \BibitemOpen
  \bibfield  {author} {\bibinfo {author} {\bibfnamefont {T.}~\bibnamefont {Odagaki}}\ and\ \bibinfo {author} {\bibfnamefont {K.~C.}\ \bibnamefont {Chang}},\ }\bibfield  {title} {\enquote {\bibinfo {title} {Real-space renormalization-group analysis of quantum percolation},}\ }\href {\doibase 10.1103/PhysRevB.30.1612} {\bibfield  {journal} {\bibinfo  {journal} {Phys. Rev. B}\ }\textbf {\bibinfo {volume} {30}},\ \bibinfo {pages} {1612--1614} (\bibinfo {year} {1984})}\BibitemShut {NoStop}%
\bibitem [{\citenamefont {Dodoo-Amoo}\ \emph {et~al.}(2013)\citenamefont {Dodoo-Amoo}, \citenamefont {Saeed}, \citenamefont {Li}, \citenamefont {Linfield}, \citenamefont {Davies},\ and\ \citenamefont {Cunningham}}]{Dodoo13}%
  \BibitemOpen
  \bibfield  {author} {\bibinfo {author} {\bibfnamefont {N~A}\ \bibnamefont {Dodoo-Amoo}}, \bibinfo {author} {\bibfnamefont {K}~\bibnamefont {Saeed}}, \bibinfo {author} {\bibfnamefont {L~H}\ \bibnamefont {Li}}, \bibinfo {author} {\bibfnamefont {E~H}\ \bibnamefont {Linfield}}, \bibinfo {author} {\bibfnamefont {A~G}\ \bibnamefont {Davies}}, \ and\ \bibinfo {author} {\bibfnamefont {J~E}\ \bibnamefont {Cunningham}},\ }\bibfield  {title} {\enquote {\bibinfo {title} {The quantum percolation model of the scaling theory of the quantum hall effect: a unifying model for plateau-to-plateau transitions},}\ }\href {\doibase 10.1088/1742-6596/456/1/012007} {\bibfield  {journal} {\bibinfo  {journal} {Journal of Physics: Conference Series}\ }\textbf {\bibinfo {volume} {456}},\ \bibinfo {pages} {012007} (\bibinfo {year} {2013})}\BibitemShut {NoStop}%
\bibitem [{\citenamefont {Lee}\ \emph {et~al.}(1993)\citenamefont {Lee}, \citenamefont {Wang},\ and\ \citenamefont {Kivelson}}]{Lee93}%
  \BibitemOpen
  \bibfield  {author} {\bibinfo {author} {\bibfnamefont {Dung-Hai}\ \bibnamefont {Lee}}, \bibinfo {author} {\bibfnamefont {Ziqiang}\ \bibnamefont {Wang}}, \ and\ \bibinfo {author} {\bibfnamefont {Steven}\ \bibnamefont {Kivelson}},\ }\bibfield  {title} {\enquote {\bibinfo {title} {Quantum percolation and plateau transitions in the quantum hall effect},}\ }\href {\doibase 10.1103/PhysRevLett.70.4130} {\bibfield  {journal} {\bibinfo  {journal} {Phys. Rev. Lett.}\ }\textbf {\bibinfo {volume} {70}},\ \bibinfo {pages} {4130--4133} (\bibinfo {year} {1993})}\BibitemShut {NoStop}%
\bibitem [{\citenamefont {Chayes}\ \emph {et~al.}(1986)\citenamefont {Chayes}, \citenamefont {Chayes}, \citenamefont {Franz}, \citenamefont {Sethna},\ and\ \citenamefont {Trugman}}]{Chayes86}%
  \BibitemOpen
  \bibfield  {author} {\bibinfo {author} {\bibfnamefont {J~T}\ \bibnamefont {Chayes}}, \bibinfo {author} {\bibfnamefont {L}~\bibnamefont {Chayes}}, \bibinfo {author} {\bibfnamefont {J~R}\ \bibnamefont {Franz}}, \bibinfo {author} {\bibfnamefont {J~P}\ \bibnamefont {Sethna}}, \ and\ \bibinfo {author} {\bibfnamefont {S~A}\ \bibnamefont {Trugman}},\ }\bibfield  {title} {\enquote {\bibinfo {title} {On the density of state for the quantum percolation problem},}\ }\href {\doibase 10.1088/0305-4470/19/18/011} {\bibfield  {journal} {\bibinfo  {journal} {Journal of Physics A: Mathematical and General}\ }\textbf {\bibinfo {volume} {19}},\ \bibinfo {pages} {L1173} (\bibinfo {year} {1986})}\BibitemShut {NoStop}%
\bibitem [{\citenamefont {Ramachandran}\ \emph {et~al.}(2017)\citenamefont {Ramachandran}, \citenamefont {Andreanov},\ and\ \citenamefont {Flach}}]{Ramachandran17}%
  \BibitemOpen
  \bibfield  {author} {\bibinfo {author} {\bibfnamefont {Ajith}\ \bibnamefont {Ramachandran}}, \bibinfo {author} {\bibfnamefont {Alexei}\ \bibnamefont {Andreanov}}, \ and\ \bibinfo {author} {\bibfnamefont {Sergej}\ \bibnamefont {Flach}},\ }\bibfield  {title} {\enquote {\bibinfo {title} {Chiral flat bands: Existence, engineering, and stability},}\ }\href {\doibase 10.1103/PhysRevB.96.161104} {\bibfield  {journal} {\bibinfo  {journal} {Phys. Rev. B}\ }\textbf {\bibinfo {volume} {96}},\ \bibinfo {pages} {161104} (\bibinfo {year} {2017})}\BibitemShut {NoStop}%
\end{thebibliography}%
\end{document}